# Outgassing properties of vacuum materials for particle accelerators

*Paolo Chiggiato*
CERN, Geneva, Switzerland

**Abstract**
Gas load and pumping determine the quality of vacuum systems. In particle accelerators, once leaks are excluded, outgassing of materials is an important source of gas together with degassing induced by particle beams. Understanding, predicting, and measuring gas release from materials in vacuum are among the fundamental tasks of ultrahigh-vacuum experts. The knowledge of outgassing phenomena is essential for the choice of materials and their treatments so that the required gas density is achieved in such demanding and expensive scientific instruments. This note provides the background to understand outgassing in vacuum and gives references for further study.

## 1. Introduction

The pressure $P$ in a given position of a vacuum vessel is obtained once the rate of gas release $Q$ and the effective pumping speed $S_{eff}$ [1] are known:

$$P = \frac{Q}{S_{eff}} + P_0 \qquad (1)$$

where $P_0$ is the ultimate pressure of the applied pumping system, i.e. the pressure ideally achieved without any gas load.

In general, in particle accelerators, the local effective pumping speed ranges between 1 and a maximum value of the order of 1000 $\ell$s$^{-1}$ due to the limited diameter of beampipes. On the other hand, the rate of release of significant gas species can extend over more than 10 orders of magnitude, roughly from 10$^{-5}$ to 10$^{-15}$ mbar $\ell$ s$^{-1}$cm$^{-2}$. Therefore, the installation of additional pumps cannot always be a remedy against the inappropriate selection of materials and their treatments. The performance of very expensive equipment could be spoiled simply because rather inexpensive solutions and guidelines are not properly adopted because of negligence or lack of knowledge.

The purpose of this note is to provide the essential background to understand spontaneous gas release from solid materials, explain the most common mitigation techniques, and introduce measurement methods.

## 2. Basic knowledge

The subject of this note is *outgassing* that is, rephrasing Paul A. Redhead [2], the spontaneous release of gas from materials in a vacuum. The deliberate removal of gas by heating or by interaction with particles, generally referred to as *degassing*, is described in detail in other CAS's contributions [3] [4].

This note does not consider sublimation as a source of gas. A vacuum expert knows that elements like Cd, Zn, Mg, Li, Ba, Pb, and Ca have vapour pressures incompatible with high-vacuum requirements and may cause severe contamination and damage to scientific apparatus. From data available in Ref. [5], the values of vapour pressures versus temperature can be plotted for all metals. Alloys containing volatile elements should be used only after a detailed investigation.

The *outgassing rate* is defined as the quantity of gas leaving a material per unit time. The letters $Q$ and $q$ represent the total outgassing rate and the outgassing rate per unit surface area (*specific outgassing rate*), respectively. Quantities of gas are generally reported as number of molecules (N), moles ($N_m$) or pressure-volume ($P \times V$). In the latter case, several units are still in practice, even though Pa.m$^3$ is the recommended one. In this note, mbar.$\ell$.s$^{-1}$.cm$^{-2}$ is generally used for outgassing rates referred to the specific surface area; Torr.$\ell$.s$^{-1}$.cm$^{-2}$ is still largely used in USA. Conversion factors can be found in Table 1.

**Table 1:** Conversion factors for different units of specific outgassing rate (T=296 K)

|  | $\dfrac{Pa \cdot m}{s}$ | $\dfrac{Torr \cdot \ell}{s \cdot cm^2}$ | $\dfrac{mbar \cdot \ell}{s \cdot cm^2}$ | $\dfrac{1}{s \cdot cm^2}$ | $\dfrac{mol}{s \cdot cm^2}$ |
|---|---|---|---|---|---|
| $1 \dfrac{Pa \cdot m}{s} =$ | 1 | 7.5×10$^{-4}$ | 10$^{-3}$ | 2.5×10$^{16}$ | 4.1×10$^{-8}$ |
| $1 \dfrac{Torr \cdot \ell}{s \cdot cm^2} =$ | 1330 | 1 | 1.33 | 3.3×10$^{19}$ | 5.5×10$^{-5}$ |
| $1 \dfrac{mbar \cdot \ell}{s \cdot cm^2} =$ | 10$^{-3}$ | 0.75 | 1 | 2.5×10$^{19}$ | 4.1×10$^{-5}$ |
| $\dfrac{1}{s \cdot cm^2} =$ | 4×10$^{-17}$ | 3×10$^{-20}$ | 4×10$^{-20}$ | 1 | 1.7×10$^{-24}$ |
| $1 \dfrac{mol}{s \cdot cm^2} =$ | 2.4×10$^7$ | 1.8×10$^4$ | 2.4×10$^4$ | 6.02×10$^{23}$ | 1 |

a) $\dfrac{Pa\, m^3}{m^2 s} = \dfrac{Pa\, m}{s}$;  b) $\dfrac{1}{s\, cm^2}$ should be see as $\dfrac{molecules}{s\, cm^2}$

Number of molecules and pressure-volume units are correlated by the ideal gas equation of state:

$$PV = Nk_B T \rightarrow N = \frac{PV}{k_B T} \qquad (2)$$

where T is the absolute temperature and $k_B$ is the Boltzmann constant. Values for the latter are:

$$k_B = 1.38 \cdot 10^{-23} \left[ \frac{N \cdot m}{K} = \frac{Pa \cdot m^3}{K} \right]$$

$$k_B = 1.38 \cdot 10^{-23} \left[ \frac{Pa \cdot m^3}{K} \right] = 1.04 \cdot 10^{-22} \left[ \frac{Torr \cdot \ell}{K} \right] = 1.38 \cdot 10^{-22} \left[ \frac{mbar \cdot \ell}{K} \right]$$

The pressure-volume units are transformed into number of molecules when divided by $k_B T$; the number of molecules per unit of pressure-volume are given in Eq. 3. Pressure-volume values are generally quoted at room temperature (T=296 K).

$$\frac{1}{k_B \cdot T} = 2.45 \cdot 10^{20} [Pa \cdot m^3]^{-1} = 3.3 \cdot 10^{19} [Torr \cdot \ell]^{-1} = 2.5 \cdot 10^{19} [mbar \cdot \ell]^{-1} \qquad (3)$$

## 2.1. Source of gas

Spontaneous gas release is due to impurities that are located on the surfaces and in the bulk of materials. An obvious source of gas is surface contamination, which is always present onto materials after manufacturing and manipulation. Typically, there are several layers of contamination (see Fig. 1). The outer part is made of oil, grease, and dirt coming from machining, manipulation, and storage in unclean rooms. Under that layer, there are molecules of different nature in direct contact with the solid. Airborne water vapour and hydrocarbons are always part of the adsorption layer. Among many other possible contaminants, halogen atoms, fluorine and sulphur are frequently detected by surface analysis methods like X-ray photoelectron spectroscopy (XPS) or Auger electron spectroscopy (AES).

Those two layers are removed chemically by solvents and/or detergents [6]. Even if a new adsorption layer is re-formed immediately after cleaning, it is supposed to be made of airborne molecules, essentially light hydrocarbons, and water vapour.

Oxide and hydroxide layers are also potential sources of gas; their typical thickness is in the order of 1 to 10 nm. Porosity in such layers can trap liquids and gas that escape in vacuum. Gas molecules can also be released from surface oxides and hydroxides, in particular during their chemical transformations.

A thick layer of highly distorted material is generally found beneath the oxide; it is referred to as 'damage skin'. Dislocations, voids, and inclusions are there in excess with respect to the deeper bulk of the material. Impurities, and most specifically gas, can be trapped in the defects and be released in vacuum.

Both oxide-hydroxide layer and damages skin can be removed by chemical etching. A new oxide layer is formed after the treatment, generally more compact. The new oxide can be stabilised by specific chemical treatments, for example by nitric acid for stainless steel or chromic acid for copper. Such treatment is called surface passivation.

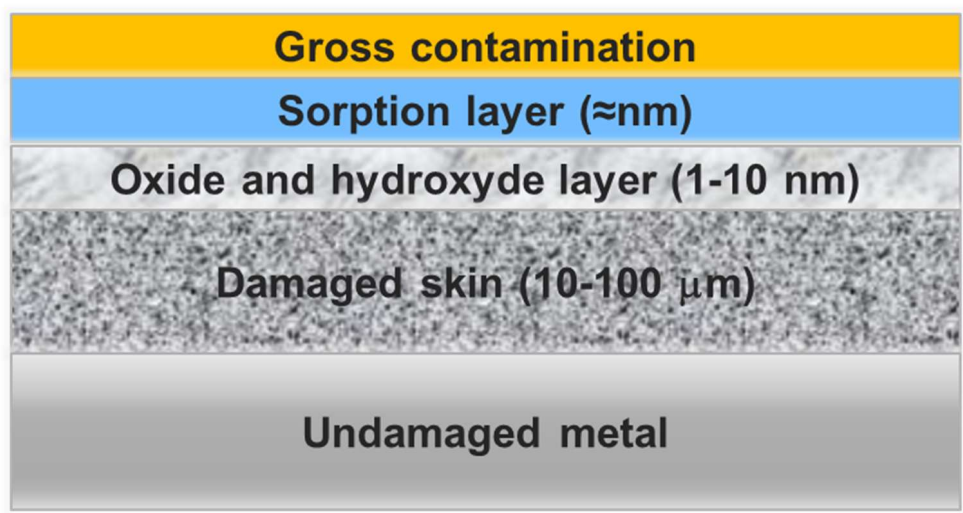

**Fig. 1:** Schematic drawing of the contamination layers [6]

The bulk of materials is also a source of gas: dissolved atoms or molecules can diffuse towards the surface. Once on the surface, such impurities can be released, possibly after chemical reaction with other adjacent molecules.

The concentration of dissolved gas considerably varies amongst materials (see Table 2). Polymers used for vacuum applications, including elastomers, can dissolve gas and liquids up to a few percent of their weight. As an example, typical water vapour content in Vespel® (a high-polyimide-based polymer as Kapton®) is 1 wt.% when it is in equilibrium with 50% humidity air at 20°C [7]. In contrast, metals dissolve only a limited quantity of small atoms; oxygen, nitrogen, carbon, and hydrogen atoms are trapped in the bulk and grain boundaries. The inclusion of these atoms occurs during melting as contaminants in ores and thermomechanical processes such as quenches and laminations. Other dissolved elements, in a lower extent unless they are intentionally added, can be halogens and sulphur. All those elements, except hydrogen, can be seen as immobile at room temperature in metals used in vacuum technology. As an example, in a day, H atoms move in average – in the meaning of diffusion length – around 4 μm in austenitic stainless steels, while for O atoms 1000 years are needed to experience the same displacement. Therefore, amongst the dissolved elements, only hydrogen is released at room temperature. For comparison, in a day, $H_2O$ molecules travel around 20 μm in PEEK (PolyEther Ether Ketone, a high mechanical performance polymer) [8].

**Table 2**: Relevant characteristics of polymers and metals for vacuum applications

| Materials | Characteristics that determine outgassing rates |
|---|---|
| **Polymers** | Dissolve entire molecules, for example $H_2O$, $O_2$, $CO_2$, solvents, etc. |
| | The solubility in the bulk can be very high in particular for $H_2O$ (up to a few weight percent, in 50% humidity air at room temperature). |
| | The dissolved molecules have a relatively high diffusivity. |
| **Metals** | Only single atoms are dissolved (H, O, C, S, etc.). |
| | The impurity concentration is in general low (typically up to $10^3$ at. ppm; for example, around 50 at. ppm for H in austenitic stainless steels). |
| | Only H has a relevant diffusivity at room temperature. |

After state-of-art cleaning [6], water vapour dominates the outgassing process in vacuum [9]. Water vapour is spontaneously released from an adsorbed layer and decomposition of hydroxides present on the surface. The reduction of $H_2O$ outgassing rate can be obtained in three different ways. Firstly, increasing the pumping time. Secondly, heating the vacuum equipment *in situ* (bakeout), so that the release rate is temporary increased, and a lower outgassing rate is achieved when the heating is completed. Finally, cooling at cryogenic temperatures, which results in lower outgassing without decreasing the number of adsorbed water molecules. The benefits obtained by the three methods are erased each time the vacuum vessel is vented to the air. This implies that pumping, bakeout or cooling are again necessary to attain the outgassing rates achieved before the venting.

Once water vapour outgassing is significantly reduced, molecular hydrogen becomes the leading outgassed species in metals [10]. Hydrogen outgassing is mainly due to spontaneous diffusion and recombination on the surface of atomic hydrogen dissolved in the bulk. In general, hydrogen outgassing rate does not depend on the duration of pumping, and an irreversible reduction of its value is obtained by heating the vacuum vessel in vacuum or in air [11].

In case of polymers, the value of outgassing rate and its trend with pumping time strongly depend on the quantity of gas dissolved in the bulk and, consequently, on the mass and geometry of the organic components. The outgassing rate can be reduced by *ex-situ* and *in-situ* thermal treatments. However, the maximum heating temperature is lower than the one for metals used in ultrahigh vacuum applications. The gas in the solid, in particular water vapour, is recharged any time the polymeric components is exposed to the air. The content of recharged water depends on temperature, humidity in air, time of exposure and, notably, the shape of the components.

The concepts introduced in the previous paragraphs are developed in the next chapters.

## 2.2. Order of magnitude of outgassing rates

Outgassing rates depend on the nature of materials and gas species, thermal and surface treatments, and temperature of operation. As it is explained in the next chapters, the time of pumping is an important parameter for polymers, and for metals as far as water vapour is concerned.

**Table 3:** Typical values of specific outgassing rates for some selected cases, measured at room temperature. N.A. means non-applicable as the outgassing rate does not depend on pumping time. The data without references were measured at CERN by Ivo Wevers, Géraldine Chuste and the author.

|  | Gas species | Pumping time [h] | q $\frac{mbar \cdot \ell}{s \cdot cm^2}$ | q $\frac{molecules}{s \cdot cm^2}$ |
|---|---|---|---|---|
| Neoprene [12] | $H_2O$ | 10 | $\approx 10^{-5}$ | $\approx 10^{14}$ |
| Unbaked austenitic stainless steel | $H_2O$ | 10 | $3 \times 10^{-10}$ | $7 \times 10^9$ |
| Unbaked austenitic stainless steel | $H_2O$ | 100 | $3 \times 10^{-11}$ | $7 \times 10^8$ |
| Austenitic stainless steel after bakeout at 150°C (24h) [13] | $H_2$ | N.A. | $3 \times 10^{-12}$ | $7 \times 10^7$ |
| OFS copper after bakeout at 200°C (24h) | $H_2$ | N.A. | $3 \times 10^{-14}$ | $7 \times 10^5$ |
| Al alloys after bakeout at 100°C-200°C (about 20h) [13] [14] [15] [16] | $H_2$ | N.A. | $\approx 10^{-13}$ | $\approx 10^6$ |
| Magnetron sputtered TiZrV thin films after activation | Kr | N.A. | $\approx 10^{-18}$ | $\approx 20$ |

As already mentioned, outgassing rates extend over several orders of magnitude (see Tab. 3). An inappropriate choice of materials and treatments leads to higher pressure than required. Most frequently, the pressure target cannot be recovered by increasing the effective pumping speed, i.e. adding new lump pumps. As an example, let us suppose a stainless-steel vacuum vessel that has an inner surface area of 1 m$^2$. Let us also suppose that a typical effective pumping speed of the order of 100 $\ell$s$^{-1}$ is applied to the vessel. After 10 h of pumping, the expected pressure is in the order of 10$^{-8}$ mbar, dominated by water vapour. If a piece of a few mm thick Neoprene of 1-cm$^2$ surface area was inserted in the vessel, i.e. only 0.01% of the total area, the same pressure would be one order of magnitude higher. Said differently, a ten times higher pumping speed would be needed to obtain the same pressure. However, the installation of additional pumps including their cabling and control system could be impossible and, even if it is possible, the additional costs could not be affordable.

As already mentioned, outgassing rates are reduced by appropriate thermal treatments. Values for H$_2$ in the 10$^{-14}$-10$^{-15}$ mbar $\ell$ s$^{-1}$cm$^{-2}$ range are achievable for metals. With such orders of magnitude, the aforementioned all-metal vessel would attain pressures in the 10$^{-12}$ mbar range only if the pumping system had an appropriate P$_o$ (see Eq. 1). As an example, turbomolecular pumps have intrinsic limitations that hinder the attainment of pressures lower than around 10$^{-10}$ mbar if not assisted by appropriate capture pumps [17].

## 3. Measurement of outgassing rate

$Q$ is measured following different procedures, depending on order of magnitude and gas species to monitor [2]. The most crucial aspect is the sensitivity of the measurement, i.e. the lowest value of $Q$ that can be reliably detected. The sensitivity is limited by the outgassing rate of the measuring system, referred to as system background $Q_R$, and the chosen type of vacuum gauges. The precision of the measurement is strongly affected by the errors of gauge calibration and, in a lower extent, by the errors in the evaluation of geometrical parameters and pumping speeds. A typical artefact that underestimates the intrinsic value of $Q$ is re-adsorption of released gas on the walls of the test system. All these aspects are treated in the next paragraphs.

### 3.1. Accumulation method

The accumulation method consists in measuring the pressure rise in an isolated vessel that contains the sample or that is itself the sample. The vessel is evacuated and, when required, baked. When the pressure is stable, the vessel is valved-off and the recording of the pressure rise starts (see Fig.2).

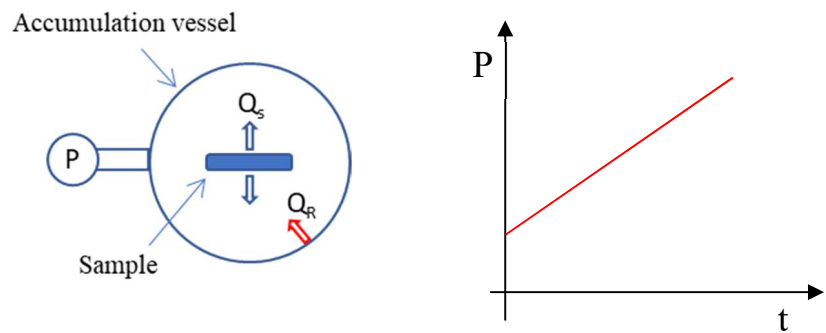

**Fig. 2:** Schematic view of the accumulation system and of a typical pressure rise of a gas that is not repumped.

If the walls of the vessel do not re-adsorb the released gas and the installed gauge has not significant pumping effects, the outgassing rate is calculated from the slope $\Delta P/\Delta t$ of the linear pressure increase. If $N$ is the quantity of gas in the accumulation volume, then

$$Q = \frac{dN}{dt} = \frac{V}{k_B T}\frac{\Delta P}{\Delta t}$$

If the quantity N is expressed in $P \times V$ units:

$$Q = V\frac{\Delta P}{\Delta t} \qquad (4)$$

which gives:

$$\Delta P = \frac{Q}{V}\Delta t \quad (5)$$

$Q$ is the sum of the gas released from the sample $Q_S$ and from the vessel including all valves, gauges and any other vacuum components exposed to the same vacuum ($Q_R$). Eq. 4 can be rewritten as:

$$Q_s = V\frac{\Delta P}{\Delta t} - Q_R \quad (6)$$

If the sample is not the vessel itself, $Q_R$ is obtained by a blank measurement that is performed without the sample, but with the same measurement procedure, namely same bakeout cycle, pumping time, measurement temperature and gauge parameters.

Residual pumping during gas accumulation results in underestimating the outgassing rates. This can be the case when the isolation valve is not fully tight, the walls of the vessel re-adsorb the released gas, or the vacuum gauge has a pumping effect. If $S_R$ is the residual pumping speed in the accumulation volume, Eq. 5 must be rewritten in the following way:

$$V\frac{dP}{dt} = Q - S_R P \quad (7)$$

where the derivative replaces the incremental ratio because the rise is not anymore linear. It gives

$$\Delta P = \frac{Q}{S_R}\left(1 - e^{-\frac{t}{\tau_p}}\right); \quad \tau_p = \frac{V}{S_R} \quad (8)$$

When $S_R$ is not zero, the pressure rises exponentially towards an asymptotic value (see Fig. 3) equal to $Q/S_R$. The error in the measurement, supposing incorrectly a linear rise, augments as the accumulation time increases.

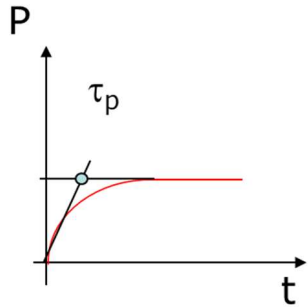

**Fig. 3:** Pressure change in an accumulation vessel where there is a residual pumping speed. The time $\tau_p = \frac{V}{S_R}$ is called the characteristic time of pumping.

The contribution to $S_R$ of leaks in the isolation valve cannot be predicted, while the one of gauges strongly depends on the measurement mechanisms. For Bayard-Alpert (BA) gauges the pumping effect is due to cracking of molecules on the hot filament and ion implantation in the collector. The pumping speeds are of the order of $10^{-1}$ $\ell s^{-1}$ for most of the relevant gas species, except noble gases that show values two order of magnitude lower [18] as the cracking mechanism is excluded. Mechanical instruments as spinning rotor [19] and capacitance diaphragm gauges have no pumping effects but requires accumulation pressure higher than $10^{-7}$ and $10^{-4}$ mbar, respectively, to provide reliable signals. The influence of surface repumping on $S_R$ depends on the gas nature. $H_2$, $CH_4$, $N_2$ and noble gases have negligible sticking probabilities on structural materials for beampipes, whereas the ones for $H_2O$, CO, $CO_2$, $O_2$ and heavier hydrocarbon are high enough to perturb the accumulation measurement.

If the accumulation volume is large enough to provide $\tau_p$ much longer than the measurement time, the value of $Q$ can be approximately calculated from the linear extrapolation of the pressure rise at the very beginning of the accumulation (McLaurin expansion of Eq. 8). For example, if $S_R \approx 10^{-1}$ $\ell s^{-1}$, $V \approx 10^4 \, \ell$, then $\tau_p \approx 10^5 \, s$, and the initial linear pressure rise is easily obtained in the first minutes of the accumulation.

For small accumulation volumes, the residual pumping precludes the use of ionisation gauges for $Q$ measurement of hydrogen, nitrogen, CO, etc. In fact, for $V \approx 10 \, \ell$ and the same residual speed as above, $\tau_p \approx 100 \, s$, i.e. after a few seconds of accumulation the pressure rise is already far from linearity. For rare gases, the gauge pumping limitation is less stringent; however, after the measurement, an ionisation gauge would not read pressures 4 orders of magnitude lower that the ones of accumulation due to the release of pumped atoms from the collector [20].

Very low hydrogen outgassing rates are measurable [21] when the accumulation volume is small (see Eq. 5) and spinning rotor gauges are used so that $S_R = 0$. As an example, for an accumulation vessel that is also the sample to measure, supposing a cylindrical shape, 2-cm radius, 20-cm length, and specific hydrogen outgassing rate of $10^{-14}$ mbar $\ell$ $s^{-1}$ $cm^{-2}$, the measurement threshold of the spinning rotor gauge is reached after three hours of accumulation; the complete measurement would last a couple of days. If the outgassing rate is 10 time lower, accumulation time around one month is required. The accumulation time can be reduced if the ratio of the sample's surface area to the accumulation volume is increased. This is achieved with bellow [21] as accumulation vessels or inserting in the volume several thin stripes. During the whole accumulation time, the temperature of the sample must be stabilized.

The accumulation method can provide an approximative time variation of the outgassing rate. During pumping, the sample is isolated, and the pressure is let increase enough to record the slope of the rise, but never more than a factor of three [2]; finally the volume is repumped. The measurement is repeated several times, and $Q$ is plotted versus pumping time. Obviously, this procedure works only if the accumulated gas does not readsorb on the surface.

### 3.2. Throughput methods

The aim of throughput methods is to measure the net gas flow removed by a pump. The sample is inserted in a vacuum vessel equipped with a gauge and a pumping system (see Fig. 4a) providing a pumping speed $S$. The gas flow is calculated by Eq. 1. In real life, this measurement is not so simple as it would appear to a superficial approach. Pressure and pumping speed must be evaluated at the right positions, avoiding pressure drops due to conductance limitations and gas beaming effects. The effective pumping speed must be calculated at the right position. To ensure uniform gas distribution, the size of the measurement vessel should be as large as possible, and the pump's opening area should be much smaller that the vessel surface area.

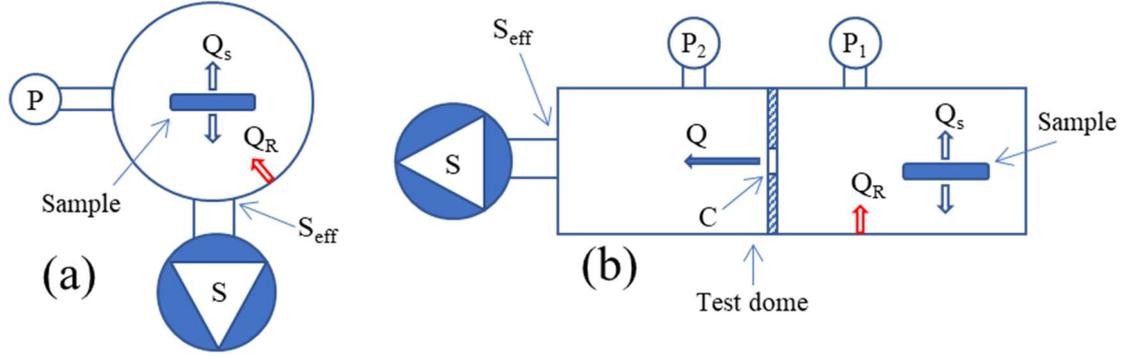

**Fig. 4:** Schematic drawings of two typical vacuum systems used for the throughput method.

To avoid a precise evaluation of $S_{eff}$, which can also depend on the pressure as it is the case of sputter ion pumps [22], another configuration is frequently used (see Fig. 4b). The gas flow is calculated by measuring the pressure drop across a known conductance $C$ applying Eq. 9.

$$Q = C(P_1 - P_2) \qquad (9)$$

$P_1$ is the pressure where the sample is located and $P_2$ is the one on the pump side. The most convenient way to have a known conductance without gas beaming is to insert a thin diaphragm with an orifice of calibrated diameter. The diaphragm can be welded on the inner surface of the measurement vessel; if made of Cu, it can also replace a CF gasket that connects the two parts of the system. It is preferable to reduce the thickness of the diaphragm around the orifice, so that gas transport is limited only by the cross-sectional area of the aperture and not by its depth. The position of the gauges is also important. As an example, Fischer and Mommsen [23] calculated the position of measurement in cylindrical domes so that Eq. 9 can be applied without error due to the anisotropy of the molecular distribution generated by the tubular dome and its orifice.

The background pressure signals ($P_{1R}$ and $P_{2R}$) are measured without the sample, while imposing the same pumping and thermal cycle of the test procedure. In this way, Eq. 9 becomes:

$$Q = C[(P_1 - P_{1R}) - (P_2 - P_{2R})] \qquad (10)$$

The recording of pressures is continuous, and it provides a time variation of Q. The pressure measurement for the throughput methods is generally provided by cold or hot ionization gauges. Their main effects on the measurement are their outgassing, which is included in the background value, and modification of the gas composition by molecular cracking on the hot filaments or by electron impact. Typical examples of cracking are $CO_2$ that transforms into CO, and $H_2$ that increases the $CH_4$ content [24]. The cracking effect is less pronounced if $C$ and $S$ are much higher than the conductance of the hot BA's filaments or the opening of the penning ionisation cell. In that cases, the molecular probability to escape the dome's volumes is much higher than the probability of cracking. In general, the pumping speeds of the ionisation gauges are not an issue in throughput methods because they are at least two orders of magnitude lower than $C$ and $S$.

Time variation of Q is obtained also for gas that are repumped on the surfaces of the vessel if the gas on the surface and in the volume are close to equilibrium, and such equilibrium is not altered by sudden pressure change. This the typical case of water vapour in metallic systems.

There are several variants of the throughput methods. The conductance modulation method [25] consists in measuring the pressure in the sample chamber with different value of C. The conductance is changed without venting the setup. For example, this can be obtained by placing a piston at different distances with respect to an orifice (see Fig. 5) or by exposing orifices of different diameter perforated in a rotary disk. For this method, the value of $S$ is not needed and only one gauge is necessary. In fact, reminding the equation for the effective pumping speed $S_{eff}$ [1]

$$\frac{1}{S_{eff}} = \frac{1}{S} + \frac{1}{C} \tag{11}$$

and using Eq.1 for two values of conductance $C_A$ and $C_B$, one obtains

$$Q = \frac{(P_A - P_{AR}) - (P_B - P_{BR})}{\frac{1}{C_A} - \frac{1}{C_B}} \tag{11}$$

where $P_A - P_{AR}$ is the pressure measured with conductance $C_A$ reduced by the background value; the same holds for index B. If $C$ can be varied in a wider range, $Q$ is obtained from the slope of the line $P$ versus $1/C$. The main disadvantage of the conductance modulation method is the outgassing rate of the device used for the change of conductance. A higher background outgassing rate limits the sensitivity of the measurement.

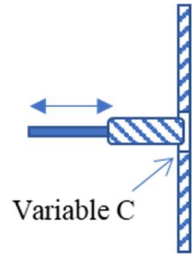

**Fig. 5:** Schematic drawings of a variable conductance: the displacement of the piston generates different geometrical restrictions to the gas flow.

To reduce the effect of the background signals and increase the sensitivity of the throughput methods, other modifications of the standard dome of Fig. 4b are reported in the literature. The two-path method is certainly worth mentioning [26] [27]. By means of two additional valves, the dome is connected to a sample holder from both side of the orifice (see Fig. 6).

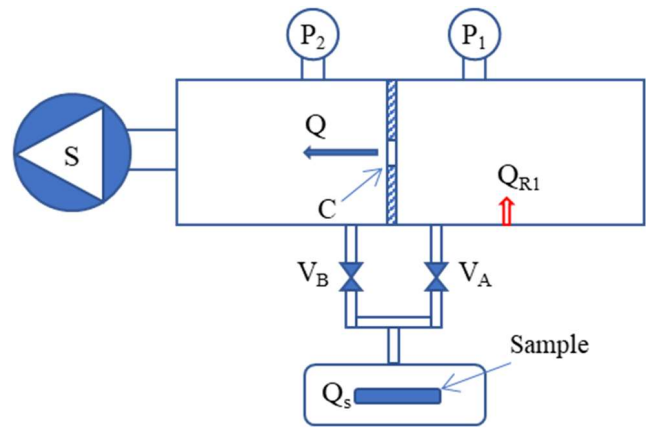

**Fig. 6:** Schematic drawings of the two-path method. The sample is inserted in an additional volume, which is connected to both sides of the orifice.

The measurement is performed in two steps with only one of the two valves open. The pressure closer to the pump ($P_2$) does not change in the two valve configurations as the gas flow is the same. Eq. 9 can be written for the two configurations:

$$\begin{cases} Q_s + Q_{R1} = C(P_{1A} - P_2) \\ Q_{R1} = C(P_{1B} - P_2) \end{cases}$$

Where the index A refers to the measurement with valve A open and B closed; the reverse is for index B. $Q_{R1}$ is the background signal due to the wall and gauge of the system located upstream of the conductance $C$. Then,

$$Q_s = C(P_{1A} - P_{1B}) \tag{12}$$

Therefore, only one gauge is strictly needed to obtain the value of $Q_s$. The drawback of this method is the added outgassing rate of the two additional valves and vacuum components to connect the sample vessel. Such an additional contribution must be evaluated by repeating the measurement without the sample. If the additional vessel is the sample, a measurement with the connecting tube sealed by blank flanges is necessary. The measurement of very low $Q$ is possible only if all components, valves included, are thoroughly degassed; as an example, in reference [26] outgassing rates of the order of $10^{-15}$ mbar $\ell$ s$^{-1}$ cm$^{-2}$ are reported.

### 3.3. Combined accumulation-throughput method

This hybrid method is frequently used at CERN for the measurement of low outgassing rate. It works only for gas that are not readsorbed on the walls of the system. The sample is isolated by a variable-leak valve in a separated vessel where the gas is accumulated for a given time $t_a$. Then the valve is progressively opened, and the quantity of accumulated gas is measured in a dome that is similar to the one of Fig. 4b, with a few modifications. The pumping speed is at least two order of magnitude higher than the conductance so that the effective pumping speed is equal to $C$. A calibrated Residual Gas Analyser (RGA) records the residual gas expansion; before installation, it is calibrated in a dedicated system, and then regularly checked *in situ* in two different ways. Firstly, a gas species is injected and the RGA reading is compared to the one of a calibrated BA gauge. Secondly, by means of a calibrated leak, a known quantity of gas is injected in the accumulation volume, and than recorded by the RGA while the variable leak valve is being opened.

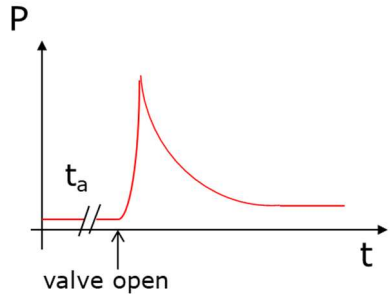

**Fig. 7:** Indicative pressure rise after the opening of the variable leak valve separating the accumulation and the measurement dome.

The value of $Q$ for the different gas species is obtained integrating the partial pressure signals from the beginning of the valve opening to the re-stabilization of the RGA signals after an interval of time $\Delta t$.

$$Q = C \cdot \int_{t_a}^{t_a + \Delta t} P(t) dt \tag{12}$$

This method can be applied only if the outgassing rate is constant in the time of accumulation and repumping is negligible. The detection limit is defined by the outgassing of the accumulation system, particularly the variable leak valve, and by the lowest signal that can be detected by the RGA. The advantages of this method are two: there are no effects of the measurement instruments on the accumulated gas, and $t_a$ can be long enough to attain RGA signals sufficiently higher than the background so that they are reliably measured.

By this method, outgassing rates of the order of $10^{-18}$ mbar $\ell$ s$^{-1}$cm$^{-2}$ were measured for Kr in a Ti-Zr-V thin-film coated vacuum chamber (standard LHC long straight sections: 7-m long, 8-cm diameter).

### 3.4. Weight loss method

The weight loss method is a standardized experimental technique largely adopted in space industry with minor applications in UHV. It is used only for materials that have a very high outgassing rate with respect to those accepted in accelerator. Typically, it is utilized to investigate the outgassing properties of organic materials, above all structural polymers, elastomers, lubricants, and adhesives. The test samples are relatively small; their dimensions are of the order of a few mm. The test starts with a day-long exposure of the samples at controlled temperature and humidity, generally 25° C and 50%, respectively. Then, after being weighed, the sample and a standardized collector plate are inserted in a vacuum system (see Fig.8) and pumped down to at least $4\times10^{-5}$ mbar. Subsequently, the sample is heated at 125°C for 24 h in front of the collector that is maintained at 25° C. After cooling at room temperature, the sample and collector are immediately reweighted. The main outcomes of the measurement are:

- the Total Mass Loss of the sample (TML);
- the weight gain of the collector (Collected Volatile Condensable Material, CVCM);
- the Water Vapour Regained (WVR) obtained after post-conditioning the sample for 24 hours in air at 20°C, 50% relative humidity.

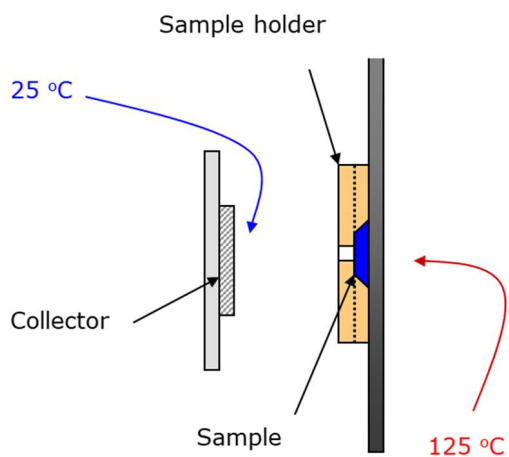

**Fig. 8:** Schematic view of the two essential elements of the weight-loss methods: the collector plate and the sample holder. The indicated temperatures are those of the standardized thermal cycle.

The set-up and procedure are described in detail with slight differences both in the ASTM E595 and ECSS-Q-ST-70-02C standards. An important collection of data is provided by NASA [7].

The throughput and weight loss methods can be coupled to obtain the kinetics of the outgassing process at room temperature and during heating at 125°C. An RGA installed in the measurement dome can provide an estimation about the gas species released by the sample. Additional information about the material condensed onto the collector is obtained by chemical analysis, for example infrared spectroscopy. In general, several samples are measured in the same run to obtain statistically relevant results. Up to 12 samples are tested at the same time by the setup available at CERN.

The weight loss method should be used only to exclude, never to accept, materials for accelerator vacuum systems. In case of non-zero CVCM, the material must be rejected to avoid risks of contamination in the beam pipes leading to multiple detrimental effects. On the other hand, a null CVCM does not mean that the material is safe in term of condensable contamination; indeed, the lowest detectable weight change by laboratory electromagnetic balance ($\approx 10^{-5}$ g) is not enough to guarantee that a negligible contamination is released in UHV systems.

## 4. Outgassing of water vapour from unbaked metal alloys

After thorough surface cleaning, water vapour dominates the outgassing process of metals in vacuum. For metal alloys, the outgassing rate is inversely proportional to the pumping time (t) and the empirical relationship of Eq. 13 is in general applied at CERN for design of vacuum systems.

$$q_{H_2O} \approx \frac{3 \times 10^{-9}}{t[h]} \left[\frac{mbar\ l}{s\ cm^2}\right] \tag{13}$$

This equation is a good approximation for all metals used in vacuum systems (stainless steel, copper, aluminium, titanium, etc.) [28] [9]. The $\approx \frac{1}{t}$ variation is valid also for a pumpdown performed at temperatures as high as 100°C [29]. This time dependence is observed in many desorption processes on solid surfaces, including many systems of interest for catalysis, and it is generally called Elovich behaviour [30] [31] [32].

The throughput methods are generally adopted for the measurement of $q_{H_2O}$, frequently in its simplest version, namely measuring pressure by a penning gauge while pumping the vessel by turbomolecular pumping systems. Measurement methods comprising accumulation are not appropriate because water is reabsorbed and, as shown in Eq. 13, there is a strong dependency of $q_{H_2O}$ on pumping time.

The total quantity of desorbed molecules is obtained by integrating Eq. 13 in the range of the pumping time. For weeks of pumping, the number of desorbed water molecules is of the order of $10^{15}$, i.e. around a monolayer [9]. Of course, the integration cannot be extended towards infinite pumping time due to the nature of the integral function ($\propto \log t$), which means that for very long pumping time $q_{H_2O}$ has to drop faster than $\frac{1}{t}$. In practice, the faster pressure drop is never measured because, soon or later, hydrogen outgassing becomes dominant. As an example, for as cleaned austenitic stainless steels, $q_{H_2O}$ and $q_{H_2}$ are comparable after 40 days of pumpdown.

It is important to stress that the outgassing rate of unbaked metals is not an intrinsic value of the material because it depends on the pumping time. Without expressively mentioning the pumping time, any value of $q_{H_2O}$ is meaningless. The pumping time dependency has a strong consequence on the operation of vacuum systems for particle accelerators. To attain a reduction of pressure by a factor of 10, the pumping time must be increased by the same factor.

There is evidence in the literature that the water vapour outgassing rate is not significantly modified by standard surface treatments, including standard electropolishing [28] . However, it has been reported that dedicated electropolishing treatments [33] are effective in reduction $q_{H_2O}$ by at least a factor of 10, while maintaining the usual $\frac{1}{t}$ decay law.

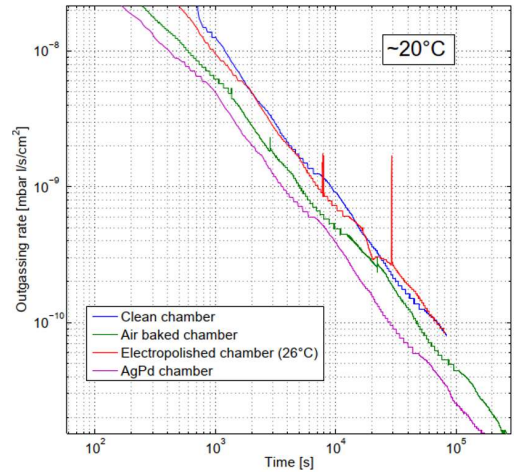

**Fig. 9:** Water vapour outgassing rates of stainless steel that underwent four different surface treatments [34] : as cleaned by detergent (blu), air baked at 300°C for 24 hours (green), electropolished (red), coated with PdAg (violet). The measurement temperature is 20°C, except for the electropolished vessel (26°C). The spikes shown in the curve of the electropolished sample are instrumental artefacts. The measurements were performed by Riccardo Renzi and José A. Ferreira Somoza at CERN.

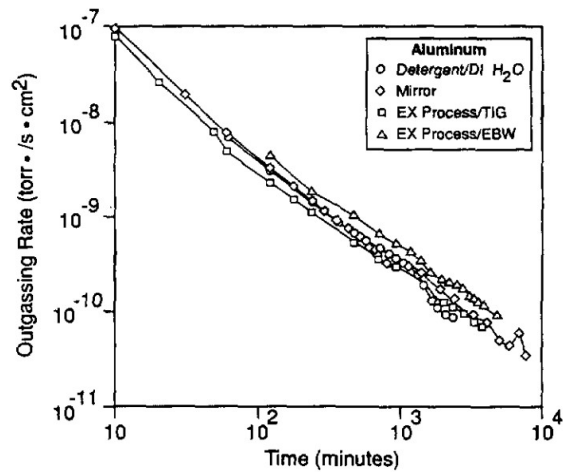

**Fig. 11:** Water vapour outgassing rate of aluminium vacuum chambers that underwent four different surface treatments [28]. One Torr is 1.33 mbar.

The characteristic decrease of $q_{H_2O}$ by pumping time is the most important obstacle in achieving low pressures in UHV systems. A faster reduction is attained by:

1. Heating the whole vacuum system during part of the pumpdown time (**bakeout**). By heating, the release of water molecules from the internal surfaces is accelerated; therefore, the pressure in the system is increased (see Fig. 12), as well as the quantity of gas evacuated. A much lower pressure is obtained when the system is cooled down to room temperature. The bakeout is very effective for metals if it is carried out for at least 12 h at temperatures higher than 120°C. Lower heating temperatures are also useful if the duration of the bakeout is significantly increased [35].

2. Permanently cooling the vacuum system to cryogenic temperatures during operation: the water molecules remain on the surfaces for a longer time, and lower pressures are quickly achieved. As an example, at 125 K the saturated vapour pressure of water is $10^{-11}$ mbar [36]. In many particle accelerators, the reduction of $q_{H_2O}$ is a bonus of cryogenic systems, which generally are necessary to attain superconductivity in magnet cables and radiofrequency cavities.

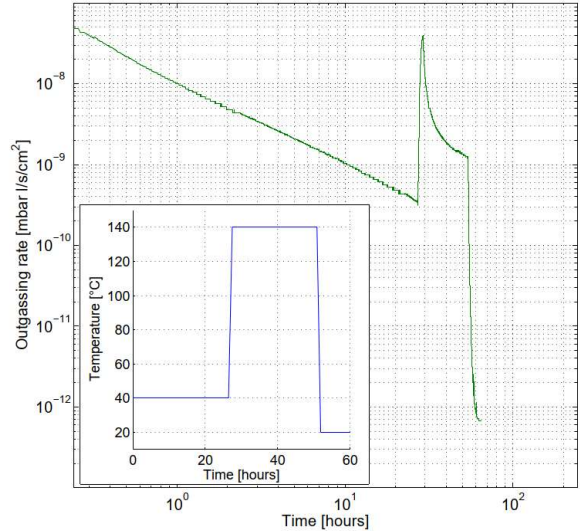

**Fig. 12:** Pressure evolution in a metallic vacuum system [34]. The pressure peak corresponds to the beginning of the bakeout (140°C). The sudden pressure drop is generated by the cooling of the vacuum vessel to room temperature. The bakeout cycle is shown in the box. The measurements were performed by Riccardo Renzi and José A. Ferreira Somoza at CERN.

In addition to bakeout and cooling, other methods are reported in the literature. For example, a gas that easily reacts chemically with water vapour can be injected in the vacuum system [37]. The new compounds resulting from the chemical reaction are either more or less volatile than water vapour. In the former case, the gas removal is faster; in the latter, the vapour pressure is reduced, and lower pressures are achieved. To my knowledge, this method is not applied to UHV systems and, by sure, not to particle accelerators. Weak points of this method are the rate of reaction and the harmfulness of the proposed gas species. In addition, to accelerate the chemical reaction, heating is recommended, which spoils the benefit of avoiding bakeout.

In another method [38] [39], UV lamps are used, in particular those based on mercury emission. The emitted radiation, in the range between 254 nm and 185 nm, is expected to excite water molecule bonding on the metallic surfaces, therefore accelerating desorption. An efficient application of UV lamps in long beam pipes of accelerators is difficult to implement.

### 4.1. Modelling the time dependency of water vapour outgassing rate

The $q_{H_2O} \propto t^{-1}$ empiric law has a major technological impact. Its understanding is important, and it depends on the nature of the interaction between water molecules and metal surfaces. One can imagine describing mathematically the time dependency assuming the simplest adsorption model, i.e. water molecules that are adsorbed on a given number of sites, $Ns$, having the same desorption energy, $E_d$. In this model, an adsorption site can either be empty or occupied by a single molecule. An adsorbed molecule can break the bonding with the surface if its kinetic energy achieves a value higher than $E_d$. The fraction of oscillations on the adsorption sites that can lead to desorption is given by the Boltzmann factor [40]:

$$e^{-\frac{E_d}{k_B T}} \tag{14}$$

The rate at which the oscillations occur can be identified as the frequency $\nu_0$ of a simple harmonic oscillator, i.e the adsorbed molecule vibrating perpendicularly to the surface. The average energy of the oscillator is $k_B T$ as given by classical mechanics. In quantum physics, the same oscillator has an energy multiple of $h\nu_0$, where $h$ is the Planck constant, i.e. 6.6×10$^{-34}$ J·s. Equating the two relations of energy, it comes out that $\nu_0 \approx k_B T/h$. Of course, this is a rough calculation that is given here only to estimate the order of magnitude of $\nu_0$. The escape frequency $\nu$ can be written as

$$\nu = \nu_0 e^{-\frac{E_d}{k_B T}} \sim \frac{k_B T}{h} e^{-\frac{E_d}{k_B T}} \sim 10^{13} e^{-\frac{E_d}{k_B T}} [s^{-1}]. \tag{15}$$

The reciprocal of the escape frequency is called the mean sojourn time $\tau$ of the molecule on the surface:

$$\tau = \tau_0 e^{\frac{E_d}{k_B T}} \sim \frac{h}{k_B T} e^{\frac{E_d}{k_B T}} \sim 10^{-13} e^{\frac{E_d}{k_B T}} [s]. \tag{16}$$

The gas release rate is calculated multiplying $\nu$ by the number of adsorption sites that are occupied. Balance equations for the number of molecules can be written for the adsorbed phase and the gas phase (all gas quantities are here reported in $P \times V$ units):

$$\begin{cases} V\dfrac{dP}{dt} = -SP + \dfrac{AN_s \theta}{\tau} \\ \dfrac{N_s d\theta(t)}{dt} = -\dfrac{N_s \theta(t)}{\tau} \end{cases} \tag{17}$$

$N_s$ and $N(t)$ are, per unit of surface area, the total number of adsorption sites and those of them that are occupied at time $t$, respectively. $A$ is the surface area.

$V\frac{dP(t)}{dt}$ is the variation of the quantity of gas in the gas phase in the unit time.

$\theta(t)$ is the fraction of adsorption sites that are occupied at time $t$: $\theta(t) = \frac{N(t)}{N_s}$

$SP(t)$ is the gas throughput at time t, assuming S constant.

$\frac{N_s \theta(t)}{\tau}$ is the gas release rate from the surface at time t.

In this very simple model, once desorbed, the water molecules are supposed to be evacuated without re-adsorption. Assuming a negligible initial pressure, for $t > \tau_P = \frac{V}{S}$, the resolution of the system leads to the following pressure variation:

$$P(t) \cong \frac{AN_s}{S \cdot \tau} e^{-\frac{t}{\tau}} \tag{18}$$

A few solutions of Eq. 17-18 are plotted in Fig. 13 for a spherical system, V=10 $\ell$ and S= 10 $\ell$ s$^{-1}$, for variable desorption energy, at room temperature. The number of sites per unit surface is assumed to be around 10$^{15}$ sites/cm$^2$, i.e. $4\times 10^{-5}$ mbar $\ell$ cm$^{-2}$. The same plots are also calculated for a temperature of 100°C.

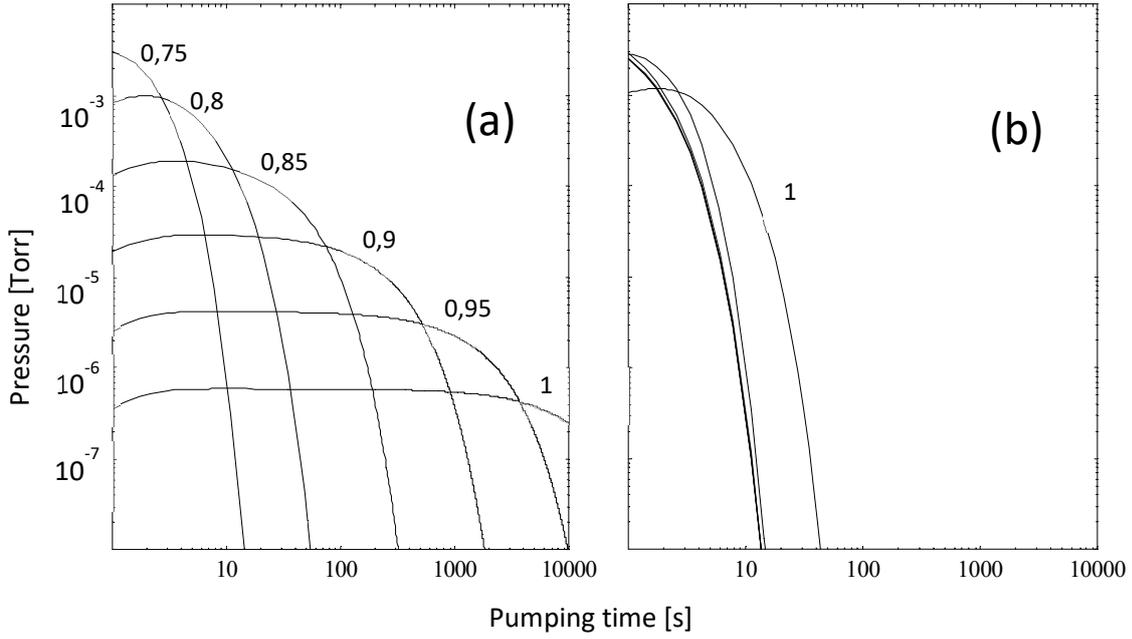

**Fig. 13:** Solutions of Eq. 17 for several desorption energies (expressed in eV) where V=10 $\ell$, S= 10 $\ell$ s$^{-1}$, $N_s$=10$^{15}$ cm$^{-2}$ at room temperature (a) and 100°C (b). The initial pressure is zero. One Torr is 1.33 mbar.

The effect of temperature is clearly shown in Fig. 13(b). This simple model predicts a beneficial effect of bakeout. However, the calculated time dependency of pressure does not fit the experimental data. This implies that the assumptions of single binding energy and no re-pumping are unrealistic. The model can be easily adapted to consider re-adsorption. For that, let us assume a constant pumping speed $S_w$ for the wall of the vessel. The system of Eq. 17 is modified by the term $S_w P$, i.e. the rate of gas repumping by the wall:

$$\begin{cases} V\dfrac{dP}{dt} = -SP + \dfrac{AN_s\theta}{\tau} - S_w P \\ \dfrac{AN_s d\theta(t)}{dt} = -\dfrac{AN_s\theta(t)}{\tau} + S_w P \end{cases} \quad (19)$$

If the characteristic time of pumping $V/S$ is smaller than the mean sojourn time, and if $S_w \gg S$ then:

$$P(t) \cong \frac{AN_s}{(S+S_w)\cdot\tau} e^{-\frac{t}{\left(\frac{S_w+S}{S}\right)\cdot\tau}} \quad (20)$$

which is again an exponential function of the pumping time. In the limit of this model, the effects of the re-adsorption is to increase the average time a molecule stays onto the walls, i.e. $\tau\left(\frac{S_w+S}{S}\right)$ instead of $\tau$, and to reduce the initial pressure.

An interesting feature appears in Fig. 13. The envelope of the exponential pressure decays calculated for different desorption energies is indeed a function of $1/t$. This means that the correct pressure decay would be reproduced if there were, at the beginning of the pumping, equally populated and independent subsets of sites with different desorption energy [41]. At each pumping time $t$, the envelop function gives the maximum attainable pressure, which is due to water desorption from sites having desorption energy close to a specific value $(E_d)_{MAX}$. This value is calculated by finding the maximum of Eq. 18 with respect to $\tau$:

$$\frac{dP}{d\tau} \cong \frac{N_s}{S\tau^2} e^{-\frac{t}{\tau}} \cdot \left(1 - \frac{t}{\tau}\right) = 0 \rightarrow \tau = t \rightarrow (E_d)_{MAX} = k_B T \ln \frac{t}{\tau_0} \qquad (21)$$

This gives desorption energies of 0.9 eV and 1.06 for 1 hour and one-month pumping time, respectively. The maximum value of $P(t)$ is [9]:

$$P(t) \cong \frac{N_s}{S \cdot \tau} e^{-\frac{t}{\tau}} = \frac{N_s}{S \cdot e \cdot t} \qquad (22)$$

and of the corresponding maximum outgassing rate:

$$q(t) \cong \frac{3 \times 10^{-5}}{e \cdot t} = \frac{1.1 \times 10^{-5}}{t[s]} = \frac{4 \times 10^{-9}}{t[h]} \left[\frac{mbar\, l}{s\, cm^2}\right] \qquad (23)$$

This value of $q(t)$ approximates in a reasonable way the empirical law of Eq. 13. The essential points behind such a formulation are that:

- A set of independent adsorption sites of different desorption energies can provide outgassing rates in agreement with the measured ones.
- Only sites with desorption energies close to a defined value ($\tau \approx t$) influence the outgassing rate at a given pumping time $t$.

### 4.2. The concept of quasi-equilibrium and the use of adsorption isotherms

A more detailed description is obtained introducing the concept of quasi-equilibrium between water arriving ($S_w P$) and leaving $\left(\frac{N_s \theta}{\tau}\right)$ the adsorbed phase on the surface of the vessel:

$$S_w P \approx \frac{AN_s \theta}{\tau} \qquad (24)$$

Assuming quasi-equilibrium and supposing $S_w \gg S$, from Eq. 19 it comes out that:

$$V \left|\frac{dP}{dt}\right| \ll S_w P \qquad (25)$$

Which gives

$$\frac{1}{P}\left|\frac{dP}{dt}\right| \ll \frac{S_w}{V} \qquad (26)$$

The meaning of Eq. 25 is that, in quasi equilibrium, the variation of the quantity of water in the gas phase must be much lower than the quantity of gas re-pumped by the surfaces per unit time. The latter inequality (Eq. 26) is referred to as Kanazawa's condition [42]; it states that in quasi-equilibrium the characteristic time of re-pumping $\left(\frac{V}{S_w}\right)$ must be much lower than the typical time of pressure variation. In other words, the adsorbed phase must always have enough time to re-equilibrate during pressure changes.

In this hypothesis, adsorption isotherms - a thermodynamic concept- can be used to correlate the coverage $\theta$ and the pressure $P$ at a given temperature T:

$$\theta = \theta(P,T)$$

Considering that

$$\frac{d\theta(P(t),T)}{dt} = \frac{\partial \theta}{\partial P}\frac{dP}{dt} \tag{27}$$

from eq. 19, summing the two equation of the system, it results that:

$$V\frac{dP}{dt} = -SP - \frac{AN_s d\theta(t)}{dt} \tag{28}$$

$$\frac{1}{P}\frac{dP}{dt} = -\frac{1}{\tau_p\left(1 + \frac{AN_s}{V}\frac{\partial \theta}{\partial P}\right)} \tag{29}$$

As already written, $\tau_P = V/S$ is the characteristic time of pumping. In case of temperature variation,

$$\frac{d\theta(P(t),T(t))}{dt} = \frac{\partial \theta}{\partial P}\frac{dP}{dt} + \frac{\partial \theta}{\partial T}\frac{dT}{dt}$$

and so

$$\frac{1}{P}\frac{dP}{dt} = -\frac{\frac{1}{\tau_p} + \frac{AN_s}{PV}\frac{\partial \theta}{\partial T}\frac{dT}{dt}}{\left(1 + \frac{AN_s}{V}\frac{\partial \theta}{\partial P}\right)} \tag{30}$$

Once the adsorption isotherm $\theta = \theta(P,T)$ is known, Eq. 29 and 30 give the pressure variation during pumpdown after resolution of the respective differential equations. However, it has to be stressed again that the quasi-equilibrium condition is not fulfilled when fast pressure changes with respect to the pumping time are imposed to a vacuum system. For example, when a valve is suddenly open or a pump is valved off after hour of pumping, the isotherm concept cannot be used anymore because the real system is too far from the thermodynamic equilibrium and the Kanazawa's condition is not valid.

### *4.2.1. The Henry isotherm*

The simplest adsorption isotherm is obtained in the limit of the infinite dilution of adsorbed water molecules on the available adsorption sites. When very few molecules are located on the surface, their presence does not significantly modify the adsorption of other molecules, which implies that the surface coverage depends only on the arrival rate on the surface, i.e. on the pressure.

$$N \propto P \rightarrow \theta_H(P) = \mathcal{H} P \tag{31}$$

where $\mathcal{H}$ is a constant of proportionality. The index 'H' of $\theta$ refers to Henry.

Eq. 31 is a different way to express the Henry law, typically used in thermodynamic to calculate vapour pressures of a solute in an infinite solution [43] [44]. Replacing Eq. 31 in Eq. 29:

$$\frac{1}{P}\frac{dP}{dt} = -\frac{1}{\tau_p\left(1 + \frac{AN_s}{V}\mathcal{H}\right)} \tag{32}$$

and integrating:

$$P(t) = P(0)e^{-\frac{t}{\left(1+\frac{AN_s}{V}\mathcal{H}\right)\cdot\tau_p}} \quad (33)$$

Therefore, the Henry isotherm gives an exponential pressure decay with a longer characteristic pumping time than a system without surface adsorption. For very low coverage and pressure, all isotherms must converge to the one of Henry. Therefore, it can be concluded that for very long pumping time, the pressure should decay exponentially. The meaning of 'very long pumping time' is not well defined unless the desorption energy distribution is known. Experience tells us that the Henry's 'regime' is not attained even after months of pumping in standard vacuum systems.

*4.2.2. The Langmuir isotherm*

In the Langmuir isotherm [45], all pumping sites have the same desorption energy $E_d$ that does not depend on coverage. The desorption rate is proportional to the filled sites, i.e. $\frac{AN_s\theta}{\tau}$, and the adsorption rate is proportional to the empty states, i.e. $(1-\theta)sPAC'$, where $s$ is the sticking probability for an empty site and $C'$ is the conductance of the unit surface area. The expression for the isotherm is obtained equating the two rates:

$$\frac{N_s\theta}{\tau} = (1-\theta)sPC'$$

which gives:

$$\theta_L(P,T) = \frac{aP}{1+aP} \quad (34)$$

and $a = \frac{sC'}{N_s}\tau = \frac{sC'\tau_0}{N_s}e^{\frac{E_d}{k_BT}}$; the index '$L$' of $\theta$ refers to Langmuir.

From equation 34 and 29, it can be obtained that:

$$\frac{1}{P}\frac{dP}{dt} = -\frac{1}{\tau_p\left(1+\frac{N_s}{V}\frac{a}{(1+aP)^2}\right)} \quad (35)$$

After integration:

$$\ln\frac{P}{P_0} + \frac{aN_s}{V}\left[\frac{1}{1+aP_0} - \frac{1}{1+aP} + \ln\left(\frac{1+aP}{P}\frac{P_0}{1+aP_0}\right)\right] = \frac{t}{\tau_p} \quad (36)$$

Fig. 14 shows the results of eq. 36 for some desorption energies. For the higher desorption energies, the calculated pressure decays display a -0.5 slope in the central part of the curve in the $\log P - \log t$ graph. This is not the behaviour that is measured. However, the envelope of the curves calculated for different energies has a -1 slope. Again, the conclusion is that different desorption energies are required to predict the real behaviour during pumpdown.

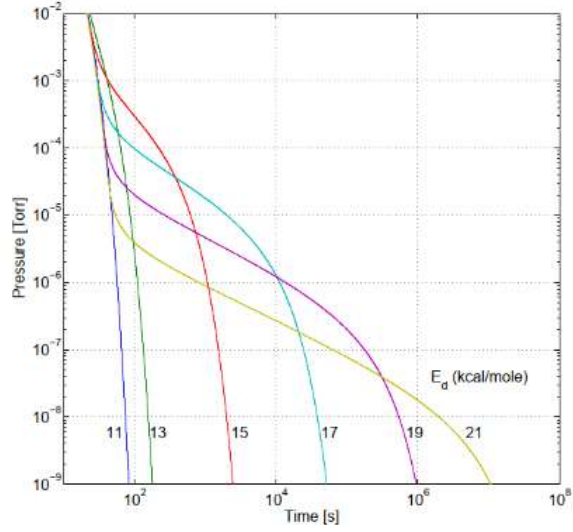

**Fig. 14:** Plots of Eq. 36 for different desorption energies [34]. The parameters are those in [34], i.e. $N_s = 3 \times 10^{15} \text{cm}^{-2}, s = 1, A = 4750 \text{ cm}^2, V = 16.7 \ \ell, S = 4.6 \ \ell \ s^{-1}$ for $N_2$. One Torr is 1.33 mbar.

### 4.2.3. General expression of adsorption isotherms

The existence of adsorption sites with different desorption energies is mathematically described by a desorption energy distribution $\varrho(E)$. The number of sites having desorption energy in the interval $[E, E + dE]$ is $N_s \varrho(E) dE$. Each sub-set of sites with energy $E$ is characterized by a 'local' adsorption isotherm $\theta(P, E, T)$ [46]. The global isotherm is given by

$$\theta(P,T) = \int_{E_{min}}^{E_{max}} \varrho(E)\theta(P,E,T)dE \qquad (37)$$

If in the interval of energy $[E, E + dE]$ the adsorption sites obeys the Langmuir hypothesis, in particular that the energy of a given site does not depend on coverage, the 'local' adsorption isotherm is $\theta_L(P, E, T)$ of Eq. 34. Eq. 37 is frequently seen as an integral equation to calculate $\varrho(E)$ when the global isotherm is measured or, the other way around, to calculate the isotherm $\theta(P, T)$ once the desorption energy distribution is known.

If all sites have the same energy $\bar{E}$; it can be written that:

$$\varrho(E) = \delta(\bar{E}) \qquad (38)$$

where $\delta(\bar{E})$ is the Dirac's delta function. The result of the integral of Eq. 37 is as expected the Langmuir isotherm for energy $\bar{E}$. The same exercise can be repeated with more realistic energy distribution $\varrho(E)$. Firstly, the adsorption isotherm is calculated with Eq. 37 supposing $\theta(P, E, T)$ equal to 'local' Langmuir isotherms; then, Eq. 29 gives the pressure variation in the quasi-equilibrium condition. Only the simple case where the density of adsorption sites is constant for each energy in the interval $[E_0, E_1]$, and zero elsewhere is detailed in this note:

$$\varrho(E) = \begin{cases} \dfrac{1}{E_0 - E_1} & \text{when } E \in [E_1, E_0] \\ 0 & \text{when } E \notin [E_1, E_0] \end{cases} \qquad (39)$$

Solving Eq. 37 with Langmuir 'local' isotherm gives:

$$\theta(P,T) = \frac{k_B T}{E_0 - E_1} \ln\left(\frac{1 + \frac{P}{P^*} e^{\frac{E_0}{k_B T}}}{1 + \frac{P}{P^*} e^{\frac{E_1}{k_B T}}}\right); \quad P^* = \frac{N_s}{C' s \tau_0} \quad (40)$$

which is the very well-known Temkin adsorption isotherm. Inserting Eq. 40 in Eq. 29, the pressure decay is obtained in the quasi-equilibrium condition.

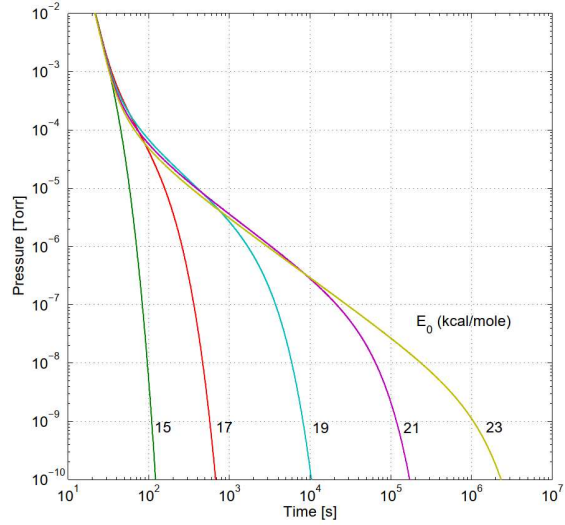

**Fig. 15:** Pressure decay calculated by Temkin adsorption isotherms having different $E_0$. The parameters are those in [47], i.e. $N_s = 2 \times 10^{16}$ cm$^{-2}$, $s = 1$, $A = 4750$ cm$^2$, $V = 16.7$ $\ell$, $S = 4.6$ $\ell s^{-1}$ for N$_2$, $E_1 = 10.6$ kcal/mol. One Torr is 1.33 mbar; 1 eV/molecule is about 23 kcal/mol.

The results are shown graphically in Fig.15. If $E_0$, the highest energy in the interval, is larger than 1 eV, the calculated curve reproduces the 1/t behaviour in a reasonable range of pumping time, i.e. from a few minutes to months. The use of the Temkin isotherm for the interpretation of water desorption in vacuum systems was already underlined at the end of the fifty by Dayton [48] and Kraus [49], and in the sixties by Schram [50].

As mentioned above, other adsorption isotherms can be used, in particular the one of Freundlich [51] [52] and Dubinin-Radushkevich (DR). The latter is considered one of the most representative isotherms for sub-monolayer gas adsorption/desorption in UHV [52]; it is extensively applied to the physisorption of gas at cryogenic temperatures [53] [54] [55]. The DR isotherm can be written as:

$$\theta(P,T) = \exp\left[-B k_B^2 T^2 \left(\ln\frac{P}{P_0}\right)^2\right] \to P = P_0 \times \exp\left[-\frac{\sqrt{\frac{|\ln \theta|}{B}}}{k_B T}\right] \quad (41)$$

$P_0$ is the saturated vapour pressure of the condensed gas at temperature T, and B is a constant. The DR isotherm is also written as:

$$\theta(P,T) = \exp\left[-\left(\frac{T}{T_0}\right)^2 \left(\ln\frac{P}{P_0}\right)^2\right] \to P = P_0 \times \exp\left[-\frac{\sqrt{|\ln \theta|}}{\left(\frac{T}{T_0}\right)}\right] \quad (42)$$

where

$$T_0^2 = \frac{1}{Bk_B^2} \tag{43}$$

Typical values of $T_0$ for water adsorption on steel are about $10^4$ K, as reported by R. Weiss [35]. It can be shown that the function $\varrho(E)$ is a skewed Gaussian [35] [52]. The DR can reasonably fit the $1/t$ behaviour of the water outgassing rate. The value of B, or $T_0$, is obtained by best fitting the experimental results. The DR isotherm was successfully applied to calculate the pressure decrease in the vacuum system of LIGO [35] . The effect of water injection at an extremity of the LIGO's arms was also calculated.

5. **Outgassing of hydrogen in metallic vacuum systems**

When water outgassing is strongly reduced, either by long pumpdown or bakeout, the outgassing process is led by $H_2$. This gas is dissolved in metals as single H atoms. Its diffusion is relatively fast and, after recombination on the surface, it is released as molecular hydrogen.

Most of the H atoms are dissolved in metals at the liquid state during the production process. Liquid metals react easily with hydrogenated molecules. Moreover, H atoms mobility and solubility in the liquid are higher than in the solid. Typical sources of H are:
- metals ores,
- tools needed for fusion,
- refractory materials of furnaces,
- combustion and treatment gas,
- water vapour and fluids used for quenching (for example the hyper-quench of austenitic stainless steels is carried out from 1100°C in water, air, or oil).

During the liquid-solid transition, H atoms are trapped in the solid at a concentration much higher than the expected equilibrium value. Residual hydrogen contents in austenitic stainless steel, copper and aluminium alloys are in the range between 0.1 and 10 ppm in weight. Higher values are reported for very reactive metals like Ti, Nb, Zr, and their alloys.

Table 4: Typical $H_2$ outgassing rates after bakeout of metals used in particle accelerators, measured at CERN by Ivo Wevers, Géraldine Chuste and the author, unless otherwise reported.

| Materials | Bakeout T[°C] for 24 h | Typical values of q [mbar $\ell$ s$^{-1}$ cm$^{-2}$] |
| --- | --- | --- |
| Austenitic st. steel | 150 | 3 10$^{-12}$ |
| Austenitic st. steel | 200 | 2 10$^{-12}$ |
| Austenitic st. steel | 300 | 5 10$^{-13}$ |
| Copper Silver added (OFS) | 150 | 3 10$^{-12}$ |
| Copper Silver added (OFS) | 200 | ≈ 10$^{-14}$ |
| Beryllium | 150 | < 10$^{-14}$ |
| AA 6060 Al alloys [56] | 150 | ≤ 10$^{-13}$ |

As for water vapour, hydrogen-outgassing rate is reduced by heating the vacuum components (see Tab. 4). Higher temperatures increase H atom mobility and, as a result, accelerate the depletion of the residual hydrogen. However, there is a crucial difference between water vapour and hydrogen. Each time the vacuum system is exposed to air, water molecules re-adsorb on the surface, while hydrogen is not recharged in the bulk of the metal. Hydrogen refilling is blocked by the difficult dissociation of the molecule on the native oxide layer. In addition, at room temperature, for most of the materials used for the manufacturing of vacuum chambers, the H solubility is very low (see section 5.1). The most important consequence of the absence of recharging is that metals keep memory of all thermal treatment they have endured. Once in air, the benefit of degassing treatments is conserved.

For copper and aluminium alloys, a few bakeout at 150-200°C for 24 hours are sufficient to reduce the hydrogen-outgassing rate to less than $10^{-13}$ mbar $\ell$ s$^{-1}$ cm$^{-2}$. For austenitic stainless steel, higher temperatures are needed to have a similar effect for a few mm thick vacuum vessel. The temperature of *in-situ* bakeout treatments are limited by thermal expansions and loss of mechanical performance of components, especially metallic gaskets. The maximum bakeout temperature of stainless-steel systems equipped with ConFlat flanges is in the range 350 to 400°C. The integrity of instruments, valves and pumps can limit even further such temperatures. This problem is circumvented by *ex-situ* treatments in vacuum or in air. A radical effect is obtained by heating austenitic stainless steels in a vacuum furnace to temperature up to about 1000°C for a few hours. Such a treatment is called 'vacuum firing'. At CERN, it is carried out at 950°C for 2 h in a vacuum better than $10^{-5}$ mbar at the highest temperature. After vacuum firing, hydrogen outgassing rates of the order of $10^{-15}$ mbar $\ell$ s$^{-1}$ cm$^{-2}$ are obtained for vacuum chambers with up to a few-mm-thick walls. These subjects are developed in detail in the next sections.

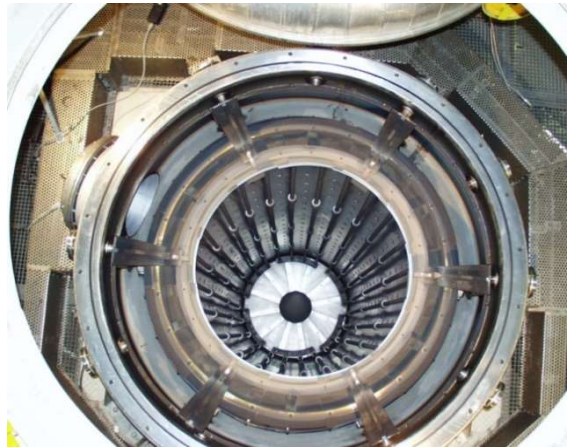

**Fig. 16**: The CERN's large furnace for vacuum firing of austenitic stainless steels. Useful height and diameter are 6 m and 1 m, respectively. Maximum charge weight: 1000 Kg. Ultimate pressure at room temperature: about $10^{-7}$ mbar; lowest pressure at 950°C: $10^{-5}$ mbar dominated by H$_2$.

## 5.1. Gas solubility and the Sieverts' law

To understand hydrogen outgassing, it is important to introduce the concept of hydrogen solubility in metals. If hydrogen in the gas phase and in the metal are in equilibrium, the concentration in the bulk $c_{eq,H}$ may be calculated by the Sieverts' law:

$$c_{eq,H}(P_{H_2}, T) = K_s(T)\sqrt{P_{H_2}} \qquad (44)$$

where

$$K_s(T) = K_0 e^{-\frac{E_s}{2k_B T}} \qquad (45)$$

$P_{H_2}$ is the pressure of hydrogen in the gas phase, frequently named 'dissociation pressure'; $K_s(T)$ is the hydrogen solubility in the metal; $E_s$ is the gas molecule solubility energy, and $K_0$ a constant. In case of endothermic reaction [57], as in most of the structural material for particle accelerators, $E_s$ is positive; therefore, the solubility increases with temperature. Eq. 46 gives the Sieverts' law for austenitic stainless steels, in different units [58].

$$c_{eq,H}(P_{H_2}, T)[H\ at.ppm] \cong 55 \cdot \sqrt{P_{H_2}[mbar]} \cdot e^{-\frac{0.115\ [eV]}{8.6 \times 10^{-5} \cdot T[K]}}$$

$$c_{eq,H}(P_{H_2}, T)[wt.ppm] \cong \sqrt{P_{H_2}[mbar]} \cdot e^{-\frac{0.115\ [eV]}{8.6 \times 10^{-5} \cdot T[K]}} \quad (46)$$

$$c_{eq,H}(P_{H_2}, T)\left[\frac{mbar\ \ell\ (H_2)}{cm^3}\right] \cong 9.5 \cdot 10^{-2} \sqrt{P_{H_2}[mbar]} \cdot e^{-\frac{2650\ [cal]}{1.99 \cdot T[K]}}$$

In such material, at room temperature and for typical hydrogen pressures encountered in vacuum technology, $c_{eq,H}$ is negligible with respect to the residual hydrogen content. For example, the hydrogen pressure in air is roughly $10^{-4}$ mbar [59], which gives an equilibrium content of around $10^{-4}$ wt. ppm. To achieve $c_{eq,H}$ of the same order of magnitude of the residual content, i.e. around 1 wt. ppm, the $H_2$ pressure must be around 7 bar. On the contrary, as a result of the endothermic nature of the reaction, at higher temperatures a significant quantity of hydrogen can be dissolved in austenitic stainless steels even for typical pressure of vacuum technology. This aspect is depicted in Fig. 17 where Eq. 46 is plotted. The line obtained by the Sieverts' law gives the equilibrium pressure as a function of temperature for 1 wt. ppm hydrogen content. Outgassing and hydrogen depletion in the material is possible only if the hydrogen pressure around the solid is kept at a value lower than the one of equilibrium. On the other hand, the material is charged in hydrogen if the pressure is higher than the equilibrium value. As written before, the plot shows that recharging at room temperature is excluded in the normal operation of vacuum systems.

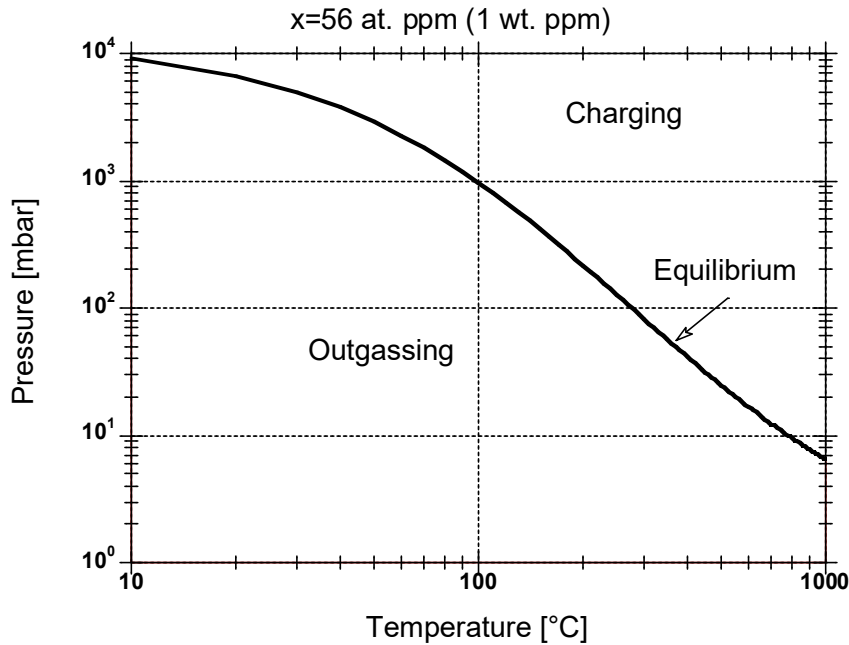

**Fig. 17**: Equilibrium pressure (dissociation pressure) of austenitic stainless steels, containing 1 wt. ppm hydrogen, at variable temperature [58]. If a pressure in the dashed area is imposed, hydrogen is released from the solid and the hydrogen content in the bulk is reduced. The opposite happens for pressure higher than the dissociation pressure.

Fig. 18 shows the dissociation pressure for different hydrogen concentrations in austenitic stainless steels. It comes out that austenitic stainless steels treated in the CERN's vacuum firing furnace cannot attain hydrogen content better than about 0.1 at. ppm. The reason is the ultimate hydrogen pressure at 950°C in the furnace (about $10^{-5}$ mbar). Since hydrogen solubility of austenitic stainless steels increases with temperature, the same content might be achieved at lower temperatures for higher residual hydrogen pressure in the furnace during the treatment, so that the vacuum performance of the furnace could be diminished. For example, if the treatment were performed at 300°C, the pressure could be 10 times higher. However, for lower temperatures or thick materials, the kinetics plays a crucial role and hinders the achievement of equilibrium. In these specific cases, hydrogen migration from the bulk to vacuum is the bottleneck of degassing. Possible metallurgical modifications must also be considered in the choice of the temperature; this includes grain growth, precipitation of chromium carbides and brittle phases, and the loss of nitrogen in nitrogen added steels.

The release of hydrogen from a solid can be concisely depicted as a sequence of two steps. In the first, hydrogen atoms randomly migrate in interstitial sites and in grain boundaries until they reach the surface. Then, the same atoms combine on the surface to form $H_2$, which is finally desorbed in the gas phase. For the sake of brevity, only the first step is considered in this note. Atomic recombination on the surface is not treated here. In the next section, it is supposed that once the H atoms reach the surface, they result in the release of molecular hydrogen without any hindrance. This hypothesis leads to results that fit well experimental data in a large range of applications in UHV technology. The reason for the good agreement might be the fact that surface diffusion is much faster than bulk diffusion, i.e. for H atoms the recombination is much faster than the time needed to reach the surface. Furthermore, for endothermic metals, the release of hydrogen molecules is not significantly limited by energy barriers [60].

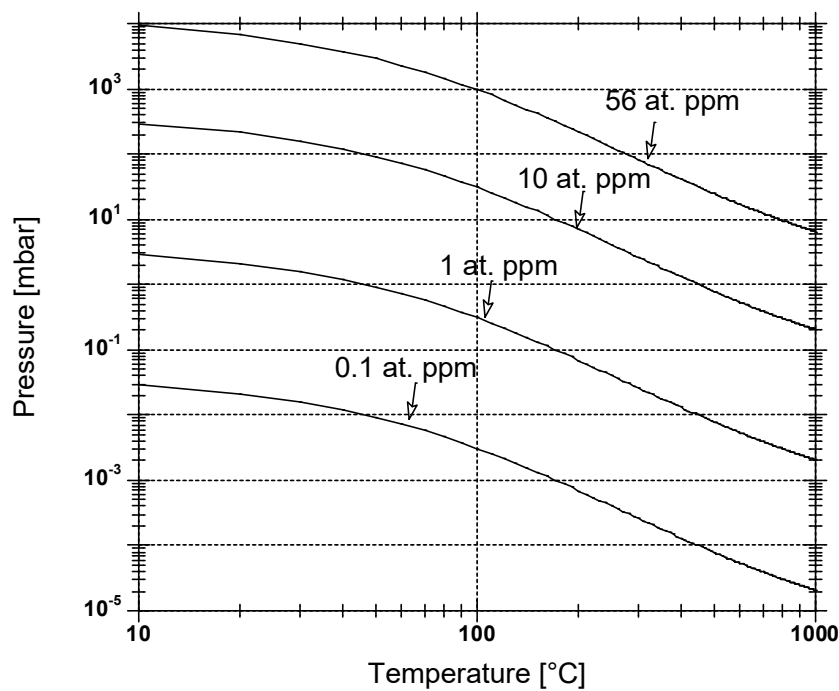

**Fig. 18**: Equilibrium pressures (dissociation pressure) of austenitic stainless steels, containing different quantities of hydrogen, at variable temperature [58]. At 950°C, a pressure in the low $10^{-5}$ mbar is in equilibrium with a solid containing 0.1 at. ppm (1.8 $10^{-3}$ wt. ppm).

## 5.2. The diffusion model of hydrogen outgassing

In the framework of a model that assumes diffusion of atomic H in the bulk as the driving phenomenon (Diffusion Limited Model, DLM), the Fick's laws should predict the measured values of $H_2$ outgassing rate. In one dimension [61], it can be written that:

$$\Gamma(x,t) = -D\frac{\partial c(x,t)}{\partial x} \tag{47}$$

$$\frac{\partial c(x,t)}{\partial t} = D\frac{\partial^2 c(x,t)}{\partial x^2} \tag{48}$$

$D$ is the diffusion coefficient of H atoms in the metals: $D = D_0 e^{-\frac{E_D}{k_B T}}$ and $E_D$ is the diffusion energy; $c(x,t)$ is the concentration and $\Gamma(x,t)$ is the flow of H atoms in the metal at position $x$ and time $t$. From the first of the two equations, the specific outgassing rate may be written as:

$$q(t) = -\frac{1}{2}D\frac{\partial c(x_s,t)}{\partial x} \tag{49}$$

$x_s$ being the ordinate at the surface. The $1/2$ factor takes into account the fact that two H atoms are needed to form an $H_2$ molecule.

To solve Eq. 47 and 48, initial and boundaries conditions must be imposed. The former, i.e. $c(x,0)$, is the initial concentration profile in the solid. For as received material, it is reasonable to assume that the concentration is roughly constant $c(x,0) = c_0$ and matches the residual content of hydrogen. The boundary condition is more subtle. As recombination and release are assumed much faster than diffusion, it is expected that hydrogen on the surface has always enough time to reach equilibrium with hydrogen in the gas phase. Strictly speaking, if the surface and the gas are perfectly in equilibrium there is no outgassing since the net flow is zero, which is the definition of equilibrium. However, it is supposed that surfaces and gas are nearly in equilibrium, so that the surface concentration $c(x_s,t)$, hereafter named $c_w$, can be approximatively evaluated by the Sieverts' law of Eq. 44 or 46:

$$c(x_s,t) = c_w \cong c_{eq,H}(P_{H_2}, T) \tag{50}$$

### 5.2.1 The semi-infinite solid approximation.

In the semi-infinite solid approximation, the concentration and the outgassing rate are:

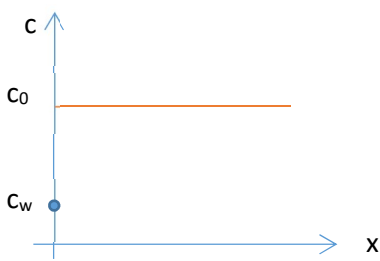

$$c(x,t) = (c_0 - c_w) \cdot erf\left(\frac{x}{2\sqrt{Dt}}\right) \tag{51}$$

$$q(t) = \frac{D(c_o - c_w)}{\sqrt{\pi D t}} \rightarrow q(t) \propto t^{-0.5} \tag{52}$$

The diffusion length $\lambda$:

$$\lambda = \sqrt{Dt} \tag{53}$$

gives an estimate of the absolute value of atomic displacement in time $t$. Eq. 52 and specifically the $t^{-0.5}$ variation of the outgassing rate are valid also for slabs when the thickness L is much larger than $\lambda$. Indeed, in this case, only a thin layer of the slab is affected by the depletion of concentration; for the rest of the material, the concentration remains almost unchanged as for the semi-infinite solid.

### 5.2.2 The slab approximation.

For a slab of thickness $L$ the boundary condition is $c\left(\pm\frac{L}{2}, t\right) = c_w$ and the initial condition, as before, is $c(x, 0) = c_0$. In this case, the calculated outgassing rate is [10] [62]:

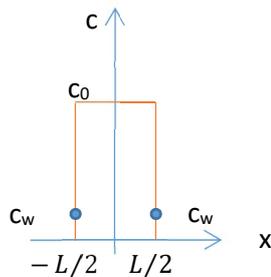

$$q(t) = \frac{4(c_0 - c_w)}{L} D \sum_{n=0}^{\infty} \exp\left[-(2n+1)^2 \pi^2 \frac{Dt}{L^2}\right] \quad (54)$$

As shown in Fig. 18, for $\boldsymbol{Dt > 0.05\, L^2}$ only the first term of the series is relevant:

$$q(t) \approx \frac{4(c_0 - c_w)}{L} D e^{-\pi^2 \frac{Dt}{L^2}} \quad (55)$$

A typical diffusion coefficient for austenitic stainless steel is given by Louthan and Derrick [63].

$$D(T)\left[\frac{cm^2}{s}\right] = 4.7 \times 10^{-3} e^{-\frac{0.56\,[eV]}{k_B T}} \quad (56)$$

It comes out that after a single 24-hour bakeout at 300°C, the condition for the validity of the Eq. 55 is fully verified for a 2-mm thick vacuum chamber.

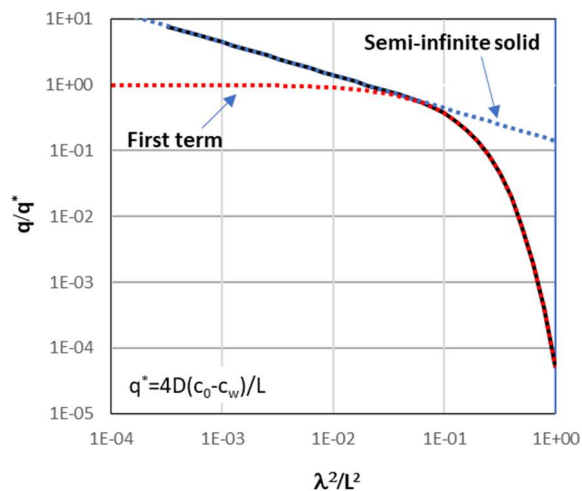

**Fig. 19:** Normalized plot of the outgassing rate given by Eq. 54 (black line) and its limits: first term only (red dashed line) and semi-infinite approximation (blue dashed line). The first term is a good approximation of the series for Dt > 0.05 L². The abscissa of the plot is the Fourier number, which is defined in Eq. 59.

### 5.2.3 Arbitrary thermal cycles

The Fick's laws (Eq. 47 and 48) can be used to calculate the outgassing rate at room temperature of slabs after arbitrary thermal cycles. In the simplest case, the material is heated at a constant temperature $T_H$ for a duration $t_H$. After cooling at room temperature $T_{RT}$, the specific outgassing rate is:

$$q(t) \approx \frac{4(c_0 - c_w)}{L} D(T_{RT}) \, exp\left[-\pi^2 \frac{D(T_H)t_H}{L^2}\right] \quad (57)$$

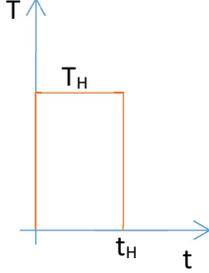

For an arbitrary temperature profile $T_H(t)$:

$$q(t) \approx \frac{4(c_0 - c_w)}{L} D(T_{RT}) \, exp\left[-\pi^2 \frac{\int_0^{t_H} D(T_H(t))\, dt}{L^2}\right] \quad (58)$$

For analogy with thermal transmission, the dimensionless number:

$$F_0 = \frac{\int_0^{t_H} D(T_H(t))dt}{L^2} = \left(\frac{\lambda}{L}\right)^2 \quad (59)$$

is called the Fourier number [64] [65]. It is the square of the ratio of the diffusion length to the slab thickness. It indicates how much of the initial residual hydrogen concentration is depleted during the thermal treatment. For $F_0>3$, the bulk of the material can be considered completely free of the residual hydrogen and in equilibrium with the surrounding gas.

In case of multiple identical thermal treatments (for example bakeout cycles) at temperature $T_H$ and duration $t_H$, the calculated ratio of the outgassing rate at room temperature after two successive thermal cycles is:

$$\frac{q_{n+1}}{q_n} = exp\left[\pi^2 \frac{D(T_H)t_H}{L^2}\right] \quad (60)$$

Therefore, each bakeout cycle reduces the outgassing rate of hydrogen by the same constant factor. This is a typical feature of the diffusion theory that can be experimentally verified (see Fig. 20 and Ref. [10]).

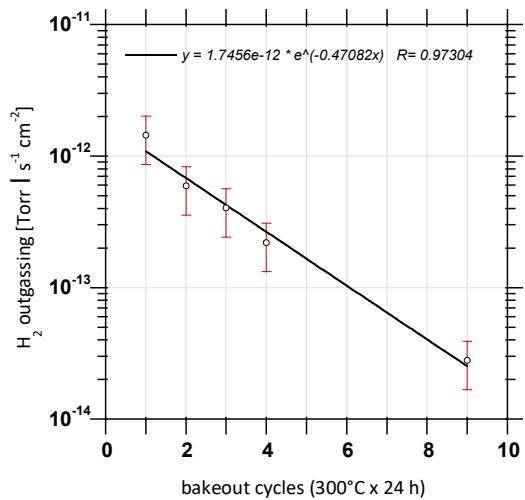

**Fig. 20**: Outgassing rate at room temperature of 316 LN sheets (2-mm thick) after bakeout cycles at 300°C for 24 hours. Each bakeout reduces the outgassing rate by a factor of about 1.6. From the exponential fit, the H diffusion coefficient at 300° is obtained, i.e. 2.2×10⁻⁸ cm²s⁻¹. From Eq. 57, an estimation of the original residual hydrogen is calculated (0.75 wt. ppm). Measurement performed by the author at CERN; 1 Torr is equal to 1.33 mbar.

### *5.2.4 Temperature dependency of outgassing rate.*

In the frame of the DLM, Eq. 54 and its approximation Eq. 58 provide the temperature dependence of the outgassing rate. If the material underwent thermal cycles at temperature much higher than $T_{RT}$, temperature excursions around room temperature does not significantly modify the exponential factor in Eq. 58; in other words, the Fourier number remains about the same. This means that the influence of temperature on $q$ can be ascribed only to the temperature dependence of the diffusion coefficient $D(T)$. The result is that the plot $\log q \propto 1/T$ is a line and the angular coefficient must be $-E_d/k_B$. This result of the DLM is easily verified experimentally (see Fig. 21).

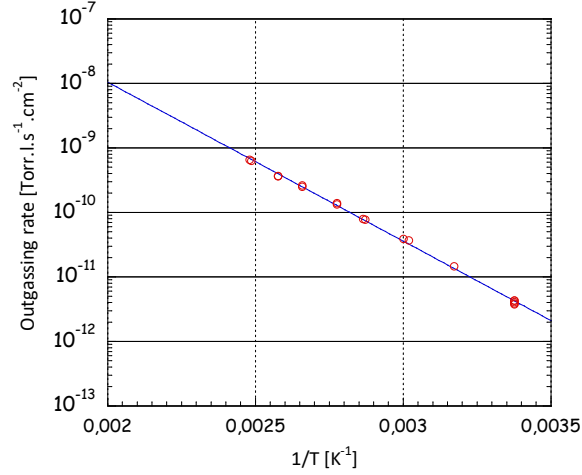

**Fig. 21**: Outgassing rate as a function of temperature measured for a 316 LN vacuum chamber (2-m long, 3.4-cm diameter, 2-mm thick) after bakeout at 200°C for 20 hours. The slope gives a value of about 0.5 eV for the diffusion energy. Measurement performed at CERN by Géraldine Chuste and the author. One Torr is 1.33 mbar.

This result is valid also for other metals used in UHV. Fig. 22 shows the case of OFS copper. The experimental results have to be compared with the values of the diffusion coefficient for copper:

$$D(T)\left[\frac{cm^2}{s}\right] = 8.34 \times 10^{-3} e^{-\frac{0.39\,[eV]}{k_B T}} \tag{61}$$

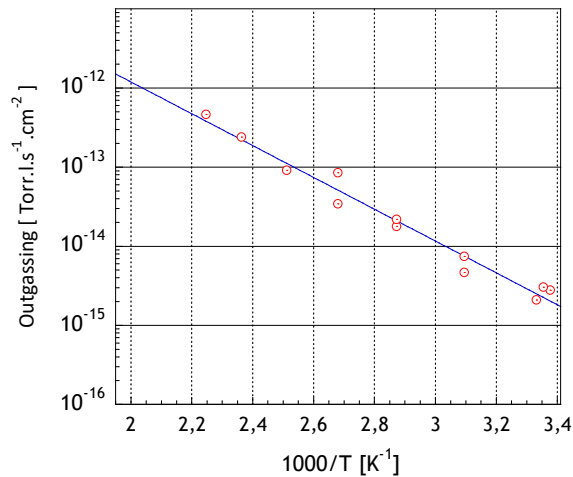

**Fig. 22**: Outgassing rate as a function of temperature measured for an OFS copper vacuum chamber (7-m long, 8-cm diameter, 2-mm thick) after bakeout at 200°C for 20 hours. The slope gives a value of about 0.4 eV for the diffusion energy. Measurement performed at CERN by Géraldine Chuste and the author. One Torr is 1.33 mbar.

## 5.2.4 The diffusion model of vacuum firing of austenitic stainless steels

As already written, vacuum firing aims to remove as much as possible hydrogen from the bulk of materials, notably austenitic stainless steels. The heating temperature is higher than that for the *in-situ* bakeout. At CERN, the standard treatment for austenitic stainless steels is 950°C for 2 hours in a vacuum of about $10^{-5}$ mbar (mainly $H_2$) at the end of the heat treatment.

For vacuum chambers thinner than 3 mm such a thermal cycle results in Fourier numbers larger than 3; that is, the residual hydrogen is completely removed, and the material reaches equilibrium with the hydrogen in the gas phase of the furnace. The hydrogen pressure during the treatment determines the final hydrogen content in the material (Sievert's law, Eq. 46) and the resulting outgassing rate at room temperature.

For sheets thicker than a few mm, the Fourier number is lower than 3, namely the diffusion length is lower than the thickness of the material $\sqrt{D_H t_H} < L$; consequently, the residual hydrogen is only partially removed. When the slab thickness is much higher than the diffusion length, the semi-infinite solid approximation can be applied. In this case, the outgassing rate does not depend significantly on the hydrogen pressure in the furnace during the treatment because $c_w \ll c_0$; the same applies to the respective dissociation pressures (see Fig. 17 that shows the hydrogen dissociation pressure for 1 ppm wt. hydrogen content at 950°C, and compare it with the typical pressure a vacuum furnace).

In real applications, after the vacuum firing ($T_f, t_f$), *in-situ* bakeout is performed ($T_{bo}, t_{bo}$) to remove water vapour. The diffusion model may be applied to calculate the resulting hydrogen outgassing rate at room temperature; Eq. 62 gives the mathematical expression.

$$q(t) = \frac{4 c_w D(T_{RT})}{L} \sum_{n=0}^{\infty} e^{-(2n+1)^2 \pi^2 \cdot F_o(T_{bo}, t_{bo})} \\ + \frac{4(c_0 - c_w) D(T_{RT})}{L} \sum_{n=0}^{\infty} e^{-(2n+1)^2 \pi^2 \cdot [F_o(T_f, t_f) + F_o(T_{bo}, t_{bo})]} \quad (62)$$

**Table 5**: Parameters of the vacuum firing treatment used for the plot of Eq. 62 in Fig. 23

| Parameters | Symbol | Values |
|---|---|---|
| Temperature of the firing treatment | $T_f$ | 950°C |
| Duration of the firing treatment | $t_f$ | 2 hours |
| In situ bakeout temperature | $T_{bo}$ | 150°C |
| Duration of the in situ bakeout | $t_{bo}$ | 24 hours |
| Initial content of residual hydrogen | $c_0$ | 1 wt. ppm wt (≈55 at. ppm) |
| Hydrogen equilibrium concentration on slab surfaces during firing | $c_w$ | 0.067 at. ppm, in equilibrium with $P_{H2}=1.3 \times 10^{-5}$ mbar |
| Density of austenitic stainless steel | | 8 g/cm³ |
| Molecular weight | | 55.845 g/mol |
| | $D_0$ | 0.0047 cm²/s |
| | $E_D$ | 0.56 eV/atom |

Fig. 23 shows the outgassing values as a function of the slab thickness for the parameters reported in Tab. 5. The thick and thin slab limits are easily identified; their outgassing rates differ by a factor of about 20. The plot shows that the hydrogen pressure in the furnace during the firing treatment plays an essential role. If the calculation did not take into account such a pressure, the outgassing rate would be largely underestimated for slab thicknesses lower than the diffusion length.

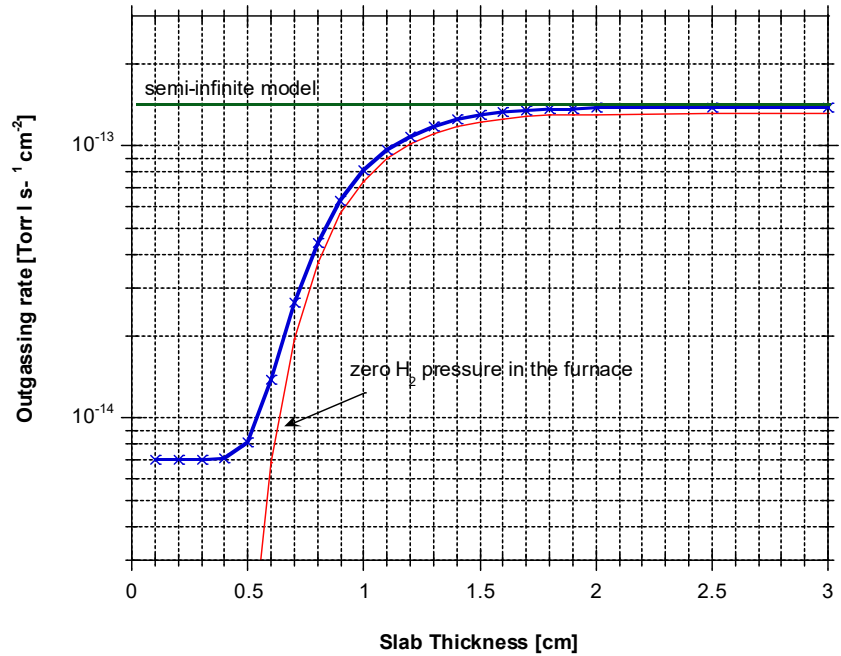

**Fig. 23**: Hydrogen outgassing rate at room temperature after vacuum firing as a function of the slab thickness (see Tab. 5 for the parameters used in the calculation). The plot shows the asymptotic value of the semi-infinite solid (green horizontal line) and the curve obtained without considering the residual pressure in the furnace (red line). The value for the lowest thicknesses depends on square root of the residual hydrogen pressure during the treatment (Eq. 44), which primarily depends on the furnace's materials and pumping system.

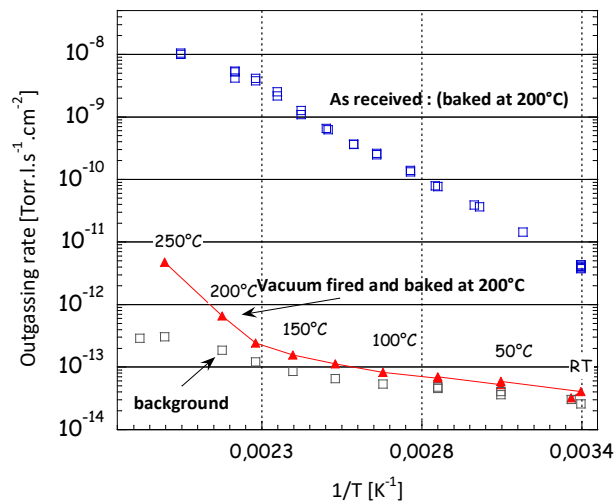

**Fig. 24**: Hydrogen outgassing rate at room temperature after bakeout at 200°C (20h) for a 316LN vacuum chamber (2-m long, 2-mm thick) as received (blue square) and vacuum fired at CERN (red triangle). The system background signal is also reported (black square); it has to be subtracted from the values of the other two curves. One Torr is 1.33 mbar.

The low values of outgassing rate after vacuum firing were verified experimentally [10] [66]; measurements performed at CERN are shown in Fig. 24. Values lower than $10^{-14}$ mbar $\ell$ s$^{-1}$ cm$^{-2}$ for a vacuum fired vessel are obtained.

*5.2.6 Limit of validity of the DLM*

The diffusion model might give wrong estimation if the slab thickness is smaller than the diffusion length and the concentrations of hydrogen in the metal is extremely low. In this case, recombination could become the limiting rate of outgassing. Shipilevsky and Glebovsky [67] calculated that a characteristic critical number of dissolved monolayers defines the limit between the two regimes; above it, diffusion is the leading process; below it, recombination becomes the controlling mechanism. For example, 1 ML of hydrogen in 1-mm thick slab of austenitic stainless steel corresponds to a concentration of the order of 0.1 at. ppm. For thicker slab, this critical concentration is reduced as the ratio of the thicknesses.

## 5.3. Air bakeout of austenitic stainless steels

Air bakeout, originally proposed by Petermann (French Patent, No 1, 405, 264) [41], is applied to austenitic stainless steels. It consists in heating vacuum components in air at temperatures in the range between 200°C and 450°C. This treatment was employed at the CERN's Intersecting Storage Rings (ISR) where up to 44 m of beam pipe were air baked at 200°C for 2 hours *in situ* [68].

The outgassing rate obtained by air bakeout is lower than the ones achieved by thermal treatment in vacuum at the same temperature and duration [69] [14]. The reason for this additional reduction is ascribed to the double effect of degassing and formation of a thicker oxide layer that acts as a barrier or a trap for hydrogen atoms. The latter effect is more relevant for air bakeout at lower temperatures, i.e. 200° to 300°C [66]. The composition of such oxide layers is predominantly iron oxide [70].

Two important applications of air bakeout are the vacuum system of LIGO and VIRGO [71] [72]. The two 3-km long arms of the VIRGO gravitational interferometer consist of 15-m long modules that are welded together (304L stainless steel, 2-mm thick, about 1.2 m diameter, corrugated walls). The air bakeout of the modules was performed at 400°C for 38 hours in a furnace. After *in-situ* bakeout at 150°C for several days, outgassing rates in the low $10^{-15}$ mbar $\ell$ s$^{-1}$ cm$^{-2}$ range were reproducibly measured.

## 6. Outgassing of polymers

In vacuum systems for particle accelerators, polymers are not used to produce vacuum vessels; the high outgassing rate and permeation are major drawbacks for this application. However, they are employed for some specific uses that encompass valve sealing, electrical insulation of cables, supports for thermal and electrical insulation, in-vacuum particle and radiation detectors, and gluing. In several applications, polymers replace ceramics and glasses, like alumina and Macor®, because they are less brittle, less expensive, and easier to machine. In UHV systems, their use should be strictly limited to unavoidable applications and only after a dedicated acceptance test. Moreover, in accelerator environment, radioactivity can induce several detrimental effects [73] [74] in the material resulting in loss of mechanical performance, brittleness, and decomposition.

## 6.1. A few notes about polymer structure

Polymers are classified according to morphology in:
- Thermoplastics: soften when heated and return to original condition when cooled (typical examples are polyethylene and PEEK).
- Thermosets: irreversibly solidify or acquires a form when heated (typical example is epoxy resin).

The polymer structure has an important influence on gas solubility and diffusion and consequently on outgassing and permeation. The two essential structures are:

- Crystalline: macromolecules form three-dimensionally ordered arrays called **lamella** (plate-like) crystals with a thickness of 10 to 20 nm in which the parallel chains are perpendicular to the face of the crystals.

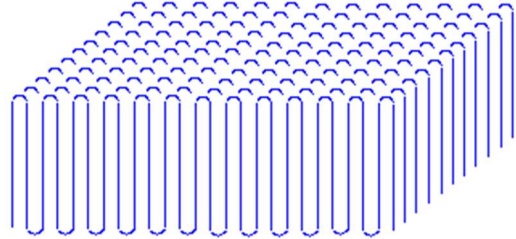

**Fig. 25**: Schematic view of a polymeric crystalline structure

- Amorphous: there is no long-range order (spaghetti like material). Amorphous polymers are softer and have lower melting points. Gas diffusion and solubility are much higher than in their crystalline phase.

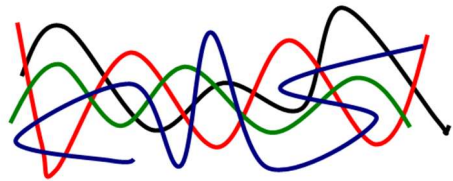

**Fig. 26**: Spaghetti-like amorphous structure of polymers

Most of the polymers contain both phases and are called semicrystalline [75]. They consist of numerous small crystallites randomly distributed through the solid and linked by interconnected amorphous areas. The degree of crystallinity is the fraction of solid composed of crystallites [76].

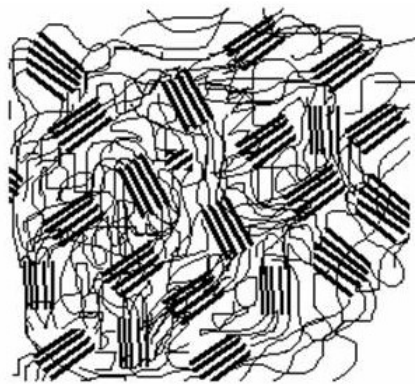

**Fig. 27**: Semicrystalline structure of polymers.

The amorphous regions can be either in the glassy or rubbery state. The temperature at which the transition from glassy to rubbery state occurs is called the glass transition temperature $T_g$. Below $T_g$, amorphous polymers are stiff, hard, and often brittle. In this state, the macromolecules are frozen in their position. Fixed micro-voids are present among the polymeric chains; they represent the so called 'excess volume'. Above $T_g$, portions of molecules start to move around; the polymer is in the rubbery state, which provides softness and flexibility. The excess volume, which is not in thermodynamic equilibrium, decreases and disappears as the temperature is raised to $T_g$. In the rubbery state, the gas molecules are dragged by the thermal movement of the chains. In the glassy state, the gas molecules diffuse through the volume of the polymer, including the micro-voids [77] [78] [79]. The excess volume plays a fundamental role in the interpretation of outgassing rate of thin polymers.

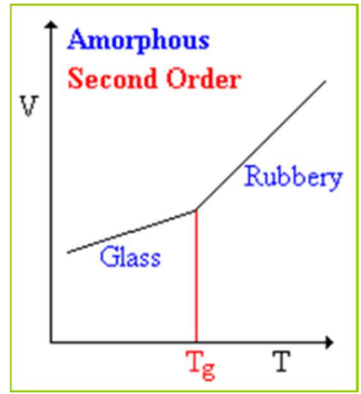

**Fig. 28**: Glassy-rubbery transition in the amorphous phase of polymers; at $T_g$ there is a second order phase transition.

## 6.2. Outgassing rates of polymers

Polymers, essentially their amorphous structures, can dissolve huge quantities of gas. Water vapour solubility is very high for UHV applications; for example, for common materials like Viton, PEEK, and Vespel® the content of water in equilibrium with 50% humidity air at 20°C is about 0.2 wt. % for the first two and 1 wt. % for the latter. The huge quantity of dissolved gas and the relatively high mobility through the polymeric chains result in much higher outgassing rate than the ones for metals. This is particularly marked for water vapour due to the high solubility.

The values of outgassing rate of polymers reported in the literature have a large spread even for material nominally identical. This could be due to:

- the large dispersion in the composition and the source of the resin,
- the different history of the samples from the liquid phase to the final component,
- the relative humidity of the laboratory,
- the shape of the component.

Data are available for baked and unbaked polymers, in particular for elastomers used for sealing [12]. The maximum bakeout temperature depends on the nature of the polymer; it is limited to about 150°C for Viton®.

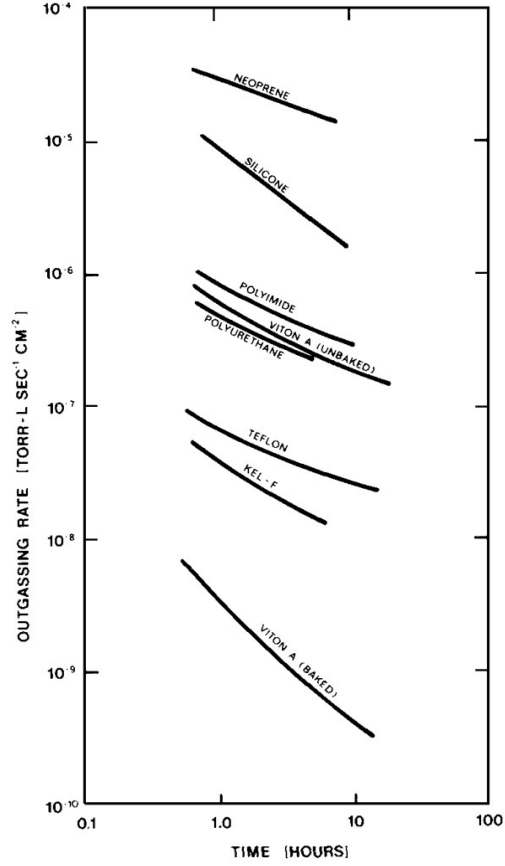

**Fig. 29:** Outgassing rates (essentially water vapour) of different polymers. The plot is copied from [12]. The data were measured by several authors and reported by R. N. Peacock in [12]. The values are reported in Torr $\ell$ s$^{-1}$ cm$^{-2}$ (1 Torr is 1.33 mbar).

Another important obstacle for the use of polymers in UHV is the high gas permeability: atmospheric gas penetrates the material and diffuses towards the vacuum system. If the vacuum system is equipped with elastomer seals, the permeation flow may limit the ultimate water vapour pressure. In addition, elastomer seals affect the sensitivity of helium leak detection because, even without any real leak, spread helium can permeate through the gaskets and be detected.

As an example, the permeation flow of atmospheric water through a Viton O-ring, 5-mm cross section diameter, 6-cm torus diameter, is about $10^{-7}$ mbar $\ell$ s$^{-1}$. The stationary condition (ultimate permeation) will be attained after about two months. Values of permeability $\Pi$ are listed in Tab. 6, where the permeation flow per cm$^2$ of exposed surface is calculated by Eq. 63.

$$q \left[\frac{mbar \cdot \ell}{s^{-1} cm^{-2}}\right] \cong 10^{-3} \cdot \left[\frac{\frac{mbar \cdot \ell}{cm^2 s} \frac{cm}{mbar}}{\frac{sccm^3}{cm^2 s} \frac{cm}{atm}}\right] \cdot \Pi \left[\frac{sccm^3}{cm^2 s} \frac{cm}{atm}\right] \cdot \frac{P[mbar]}{L[cm]} \qquad (63)$$

where $L$ is the thickness of the polymeric gasket (for example, the torus diameter of a gasket) and P the pressure of the permeating gas in air (for example 20 mbar of water vapour in air).

**Table 6:** Room temperature (20-30°C) permeability of different molecules in polymers. The table is copied form reference [12]. The data reported by R. N. Peacock in [12] were measured by several authors. For a detail description of the measurement procedures, the reader should refer to the references reported in [12]. Values are in sccm s$^{-1}$ cm$^{-2}$ cm atm$^{-1}$ (sccm is standard cubic centimeter; atm is atmosphere).

| Polymer | Helium (K × 10$^8$) | Nitrogen (K × 10$^8$) | Oxygen (K × 10$^8$) | Carbon dioxide (K × 10$^8$) | Water (K × 10$^8$) |
|---|---|---|---|---|---|
| Fluoroelastomer | 9–16 | 0.05–0.3 | 1.0–1.1 | 5.8–6.0 | 40 |
| Buna-N | 5.2–6 | 0.2–2.0 | 0.7–6.0 | 5.7–48 | 760 |
| Buna-S | 18 | 4.8–5 | 13 | 94 | 1800 |
| Neoprene | 10–11 | 0.8–1.2 | 3–4 | 19–20 | 1400 |
| Butyl | 5.2–8 | 0.24–0.35 | 1.0–1.3 | 4–5.2 | 30–150 |
| Polyurethane | --- | 0.4–1.1 | 1.1–3.6 | 10–30 | 260–9500 |
| Propyl | — | 7 | 20 | 90 | — |
| Silicone | — | — | 76–460 | 460–2300 | 8000 |
| TEFLON | — | 0.14 | 0.04 | 0.12 | 27 |
| KEL-F | — | 0.004–0.3 | 0.02–0.7 | 0.04–1 | — |
| Polyimide | 1.9 | 0.03 | 0.1 | 0.2 | --- |

### 6.3. An example: outgassing rate of PEEK disks

A typical measurement of outgassing rate is reported here. It concerns 2-mm thick disks made of PEEK. Part of the sample was cut in 200-300 mg pieces and underwent a weight loss measurement as presented in chapter 3.4. The results for the TLM and CVCM were 0.21 % and 0.01%, respectively. The latter value is not significant as it is of the order of the microbalance sensitivity.

The measurement of the outgassing rate was performed by the throughput method; at the same time, an RGA recorded the nature of the released gas. The outgassing rate depends on pumping time as shown in Fig. 30. After 10 hours of pumping the outgassing rate is 500 times higher than the equivalent one for metals. The main gas is by far water vapour.

After about 3 hours of pumping, the slop in the $\log q \propto \log t$ plot is -0.5. This indicate a diffusive process described by the semi-infinite model. As the TLM is mostly due to water, the value of $c_0$ is known and Eq. 52 gives the diffusion coefficient by best fit; $c_w$ is fixed to zero as it is much smaller than $c_0$. The obtained value is

$$D_{H_2O} = 3.9 \times 10^{-9} \left[\frac{cm^2}{s}\right]$$

that is comparable to a literature value of $5 \times 10^{-9}$ cm$^2$s$^{-1}$ [8].

Then, another similar disk was measured by the combined accumulation-throughput method described in chapter 3.3. The purpose was to measure outgassing rates after bakeout at 125° for 20 hours. The RGA detected that water vapour was the leading gas even after bakeout; as shown in Fig. 31, peak 18 amu was more than a factor of 10 higher than the peaks of masses 2, 28 and 44 amu.

As already explained, the accumulation methods cannot provide valid results for the outgassing rate of water. This is especially true when measuring polymers. A rough estimation gave a value in the low 10$^{-8}$ mbar $\ell$ s$^{-1}$ cm$^{-2}$. In addition, due to the very high partial pressure of water, hydrogen outgassing could not be measured as peak 2 amu was mainly due to the fragmentation pattern of the H$_2$O molecule. The measured outgassing rate of N$_2$, CO$_2$, and Ar are $2.5 \times 10^{-10}$, $1.3 \times 10^{-10}$, and $8 \times 10^{-1}$ mbar $\ell$ s$^{-1}$ cm$^{-2}$, respectively. The measurement for Ar is shown in Fig. 32.

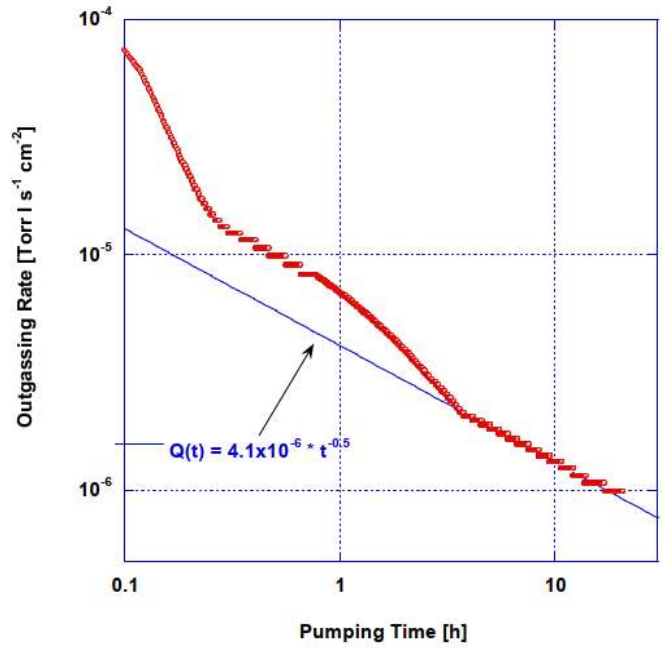

**Fig. 30:** Outgassing rate evolution with pumping time of a 2-mm thick PEEK disk. The contribution of the system background is negligible. After 3 hours of pumping, the best fit is given by the inverse of the square root of the pumping time. (1 Torr is 1.33 mbar).

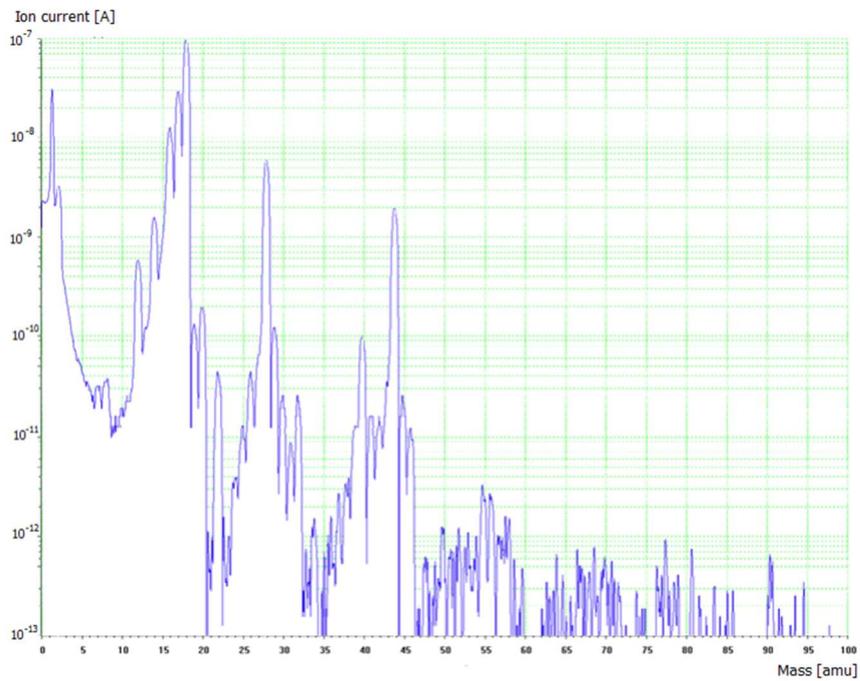

**Fig. 31:** RGA scan from mass 0 to 100 amu after bakeout at 125°C for 20 hours. The main gas released from the PEEK sample is water (18 amu). The ratio of the RGA signals for peak 28 amu to the one of 14 amu indicates that the contribution to the former is mainly due to $N_2$.

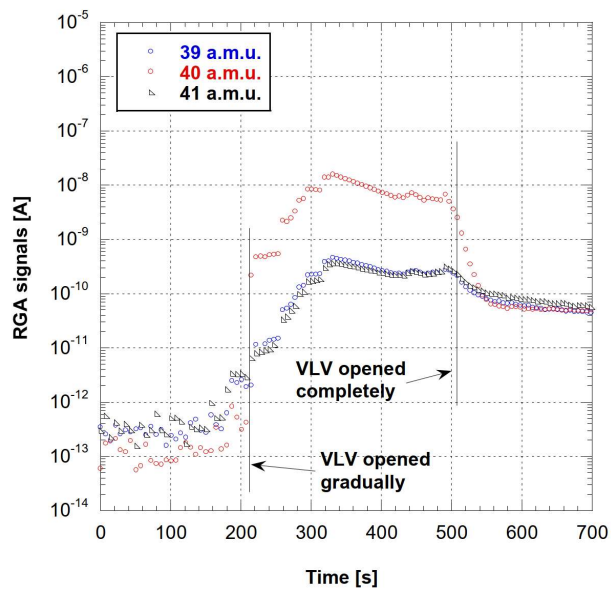

**Fig. 32:** RGA signal for Ar (mass 40 amu) during the release of the accumulated gas (accumulation time is 16 hours). The arrows indicate the time when the variable leak valve (VLV) starts to be opened and when it is completely open. The nearby peaks 39 and 41 are shown to ensure that Ar, and not $C_3H_8$, is the real source of the peak 40 amu.

The accumulation measurement was repeated at higher temperatures. The data were reported in a typical Arrhenius plot ($\log q \propto 1/T$) to obtain activation energy for the outgassing process (see Fig. 33). The resulting values are 0.33, 0.34, and 0.27 ev/molecule for $N_2$, $CO_2$, and Ar, respectively.

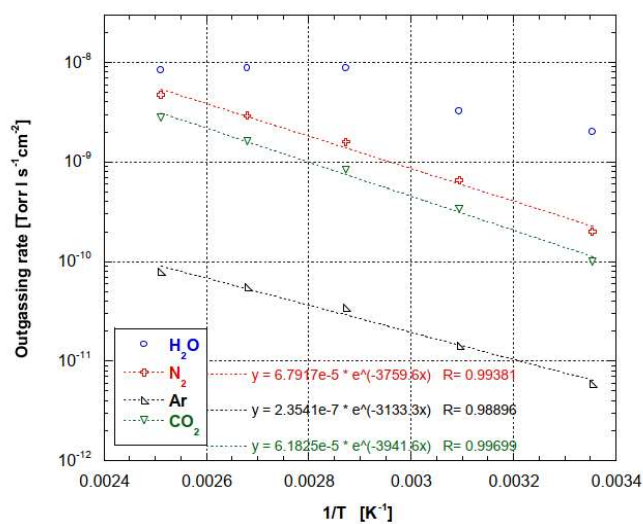

**Fig. 33:** Arrhenius plot for the outgassing rates of $N_2$, Ar and $CO_2$. The accumulations lasted about 6 hours at each measurement temperature. The data for water vapour are also plotted but not fitted as the measurement is affected by readsorption.

### 6.4. Another example: outgassing of thin Kapton HN foils.

The outgassing rates of polymeric foils of thickness within the range between 0.0125 and 0.125 mm were measured in the frame of a systematic study reported in Ref. [80]. The data for Kapton HN are shown in Fig. 34. This material is frequently used in vacuum technology for its electrical and thermal insulation properties, and its resistance to radiation.

The measurement was performed by the throughput method. As expected, water was the leading gas. For short pumping time, the outgassing rate of all samples decreased following the law $(t) \propto 1/\sqrt{t}$; then, the decrease rate was much faster. The time at which the change of regime happened increased with the thickness of the sample.

The results were initially interpreted by the diffusion theory. The slab approximation was applied. For short pumping time the relation $(\lambda/L)^2 < 0.05$ is valid and the semi-infinite model can be used to fit the initial data (see Eq.52). Then, when $(\lambda/L)^2 > 0.05$, the outgassing rate decays exponentially as shown in Eq.55. However, as clearly seen in Fig. 33, the tails of the curves are not exponential; their trend is most likely to be $q(t) \propto t^{-3}$. The desorption is delayed with respect to the expected exponential behaviour as if an additional source of gas played a leading role in that phase of the pumpdown. This was attributed to water molecules dissolved in the micro-voids of the amorphous phase.

A global fit of the curves gave a water diffusion coefficient of about $1.7 \times 10^{-9}$ cm$^2$ s$^{-1}$ and an initial content $c_0$ around 1 wt. %.

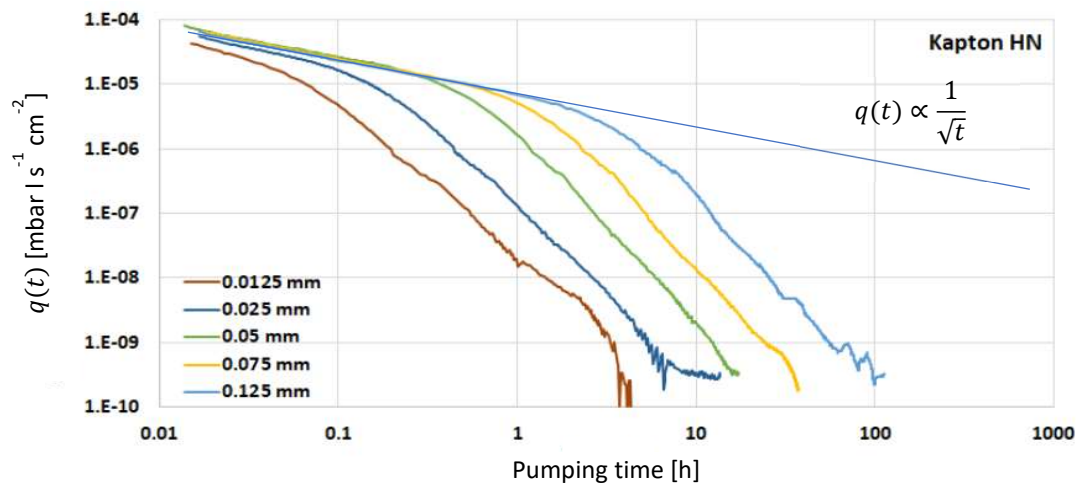

**Fig. 34**: Outgassing rates $q(t)$ of Kapton foils of different thickness. The straight line represents the $q(t) \propto 1/\sqrt{t}$ behaviour. The background value of the experimental setup was removed. Sergio Giacomo Sammartano, Ivo Wevers and Giuseppe Bregliozzi performed the measurements at CERN. More results are available in Ref. [80].

### 6.5. Reduction of the outgassing rates of polymers by thin film coating

The drawback of high outgassing of polymers has been partially overcome in other technological domains (for example packaging) by coating with thin films of metals, metal oxides, and more recently a:C-H [81] [82] [83]. Since the permeability in metals of all gas species, except hydrogen, is negligible, such coatings should entirely block the outgassing from the bulk. However, experimental results show that only a partial reduction of the gas flow is obtained. This is attributed to defects in the coating (pinholes or scratches) that cause discontinuity in the impermeable layer [81]. Pinholes are generated during the deposition process and they are presumably due to atmospheric dust particles.

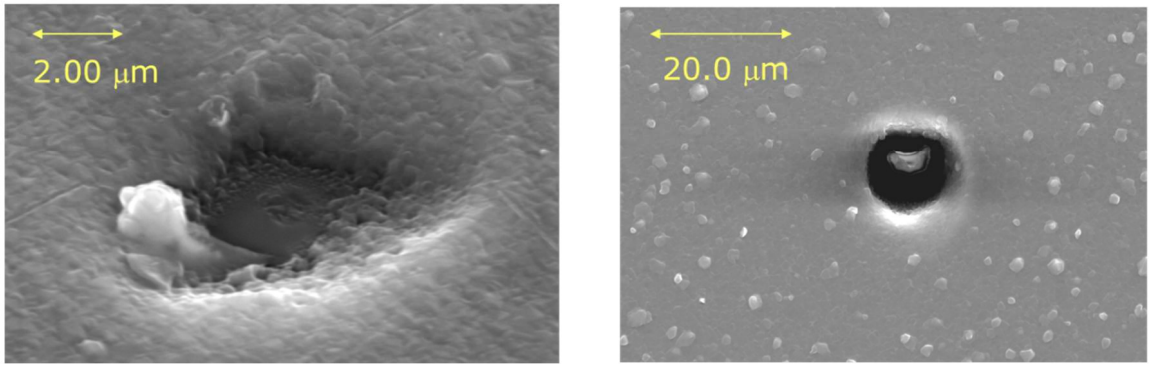

**Fig 35:** Examples of pinholes on a 2-μm thick Al coating deposited on PEEK [84]

Indeed transmitted light optical microscopy has confirmed that the uncoated part of surfaces consists of pinholes in the film [81] (see Fig. 35). The fraction of uncoated surface θ for Al coating deposited on polymeric substrates treated in laboratory air are of the order of $10^{-4}$ [84]; lower values should be attained if the substrates are treated in clean rooms. The distribution of the pinhole size is shown in Fig. 36.

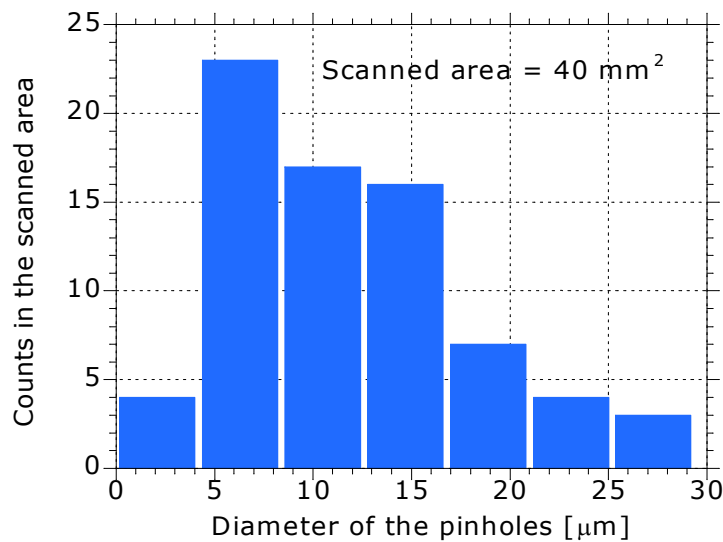

**Fig. 36:** Distribution of pinholes diameters measured by transmitted light and optical microscopy on a Mylar foil coated with a 2-μm thick Al. The measurement was performed by Cristina Bellachioma and it is explained in detail in Ref. [84].

The permeability of coated polymers is reduced much less than expected from the uncoated fraction. For example, for Al coated PEEK (0.125 mm thick), the uncoated fraction is $10^{-4}$, but the permeability of helium is only $10^{-2}$ of the one of the uncoated polymers [84] [85]. This apparent inconsistency can be justified considering that the gas flow through the pinhole is enhanced by lateral diffusion (see Fig. 37).

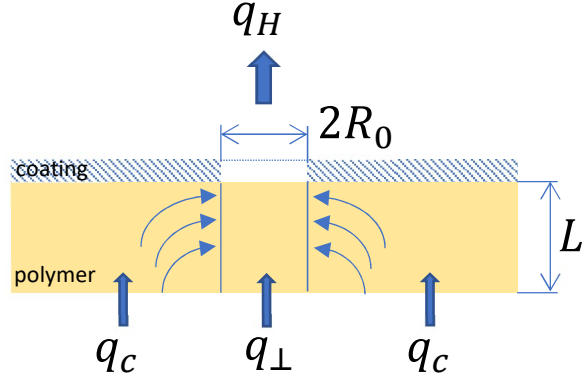

**Fig. 37:** Pinhole flow ($q_H$) enhancement due to lateral contribution ($q_c$). $q_\perp$ is the net diffusion flow that would pass through the pinhole without lateral contribution. $R_0$ is the radius of the pinhole and $L$ is the thickness of the polymer.

It can be shown [85] that for $L/R_0 > 0.3$ the normalized permeability ρ, i.e. the ratio of the coated to the uncoated substrate permeabilities, is given by:

$$\rho = \Theta \times \left(1 + 1.18 \frac{L}{R_0}\right) \tag{64}$$

$R_0$ is the radius of the pinhole. In the literature the barrier efficiency of the coating is called "barrier improvement factor" or BIF. It is the inverse of normalized permeability.

$$BIF = \frac{1}{\rho} = \frac{1}{\Theta \times \left(1 + 1.18 \frac{L}{R_0}\right)} \tag{65}$$

Eq. 65 shows that, if the size of the pinhole is constant, the improvement due to the coating is less significant for thicker polymer substrates. The BIF values are shown in Fig. 38 for the case reported in Ref. [84].

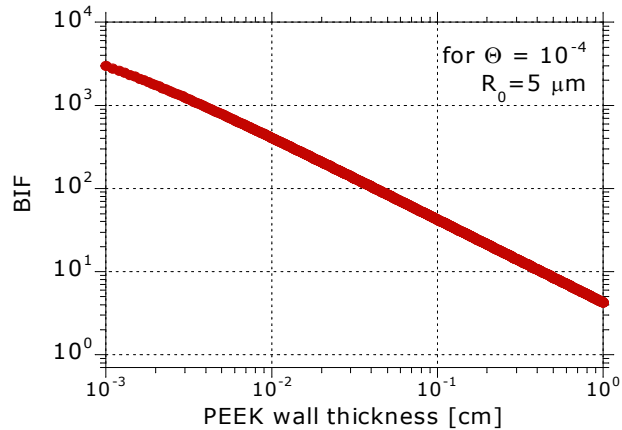

**Fig. 38:** Barrier improvement factor plotted as a function of the thickness of the substrate. The uncoated fraction and the pinhole radius are those of Ref. [84].

Metallic coatings are very efficient in the reduction of the permeation and outgassing flow in thin polymers. For polymers thicker than 5 mm, the benefit of the coating is less than a factor of 10. Better results should be achieved if the polymer is treated in a clean room.

## 7. Conclusions

The main topics concerning outgassing of vacuum materials for particle accelerators have been introduced in the previous chapters. The quality of a vacuum system is firstly based on the choice of materials and their treatments. Solubility, diffusion, and surface adsorption are among the essential properties that determine the outgassing of materials. Even though this note gives details in specific subjects, it is far to provide a complete view. Families of materials are not covered, as ceramics, glasses, and graphite. Specific treatments are not mentioned, as plasma treatments, mechanical and laser surface modifications. If there is a message that should be retained from this note, it is that measurement of outgassing is essential for any material of interest. A vacuum expert for particle accelerators is first and foremost an outgassing measurement specialist.

### References

[1] P. Chiggiato, "Vacuum Technology for Ion Sources," in *CERN Accelerator School, CAS, 2013*, https://arxiv.org/abs/1404.0960.

[2] P. A. Redhead, "Recommended practices for measuring and reporting outgassing data," *J. Vac. Sci. Technol. A,* vol. 20, no. 5, pp. 1667-1675, 2002.

[3] O. Gröbner, "Dynamic outgassing," in *CERN Accelerator School, CAS, Vacuum Technology, CERN-99-05, 1999*, http://cds.cern.ch/record/402784/files/CERN-99-05.pdf.

[4] N. Hilleret, "Non-thermal outgassing," in *CERN Accelerator School, CAS, Vacuum in accelerators, CERN-2007-003, 2007*, http://cds.cern.ch/record/923393/files/CERN-2007-003.pdf.

[5] "Vapor Pressure Calculator," Institut für Angewandte Physik - Technischen Universitat Wien, [Online]. Available: https://www.iap.tuwien.ac.at/www/surface/vapor_pressure. [Accessed 2020].

[6] M. Taborelli, "Cleaning and surface properties," in *CERN Accelerator School, CAS, Vacuum in accelerators, CERN-2007-003*, http://cds.cern.ch/record/923393/files/CERN-2007-003.pdf

[7] "Outgassing Data for Selecting Spacecraft Materials," NASA, 2018. [Online]. Available: https://outgassing.nasa.gov/. [Accessed 2020].

[8] G. Mensitieri, A. Apicella, J. M. Kenny and L. Nicolais, "Water sorption kinetics in poly(aryl ether ether ketone)," *J. of Applied Polymer Science,* vol. 37, p. 381, 1989.

[9] D. Edwards Jr., "Upper bound to the pressure in an elementary vacuum system," *J. Vac. Sci. Technol.,* vol. 14, no. 1, pp. 606-610, 1977.

[10] R. Calder and G. Lewin, "Reduction of stainless-steel outgassing in ultra-high vacuum," *Brit. J. Appl. Phys.,* vol. 18, pp. 1459-1472, 1967.

[11] G. Lewin, Fundamentals of vacuum science and technology, McGraw-Hill, 1965.

[12] R. N. Peacock, "Practical selection of elestomer materials for vacuum seals," *J. Vac. Sci. Technol.,* vol. 17, no. 1, p. 330, 1980.

[13] C. Benvenuti, "Extreme Vacua: Achievements and Expectations," *Physica Scripta,* vol. T22, pp. 48-54, 1988.

[14] G. Moraw and R. Dobrozemsky, "Attainment of Outgassing Rates Below 10E-13 Torr liters/sec cm^2 for Aluminium and Stainless Steel After Bakeout at Moderate Temperatures," *Jpn. J. Appl. Phys.,* vol. 13, pp. 264-264, 1974.

[15] H. J. Halama, "Performance of vacuum components operating at 1x10E-11 Torr," *J. Vac. Sci. Technol.,* vol. 16, no. 2, pp. 717-719, 1979.

[16] M. Suemitsu, T. Kaneko and N. Miyamoto, "Aluminium alloy ultrahigh vacuumm chamber for molecular beam epitaxy," *J. Vac. Sci. Technol. A,* vol. 5, no. 1, pp. 37-43, 1987.

[17] K. Jousten, Handbook of Vacuum Technology, 1 ed., Wiley-VCH, 2008, pp. 441-442.

[18] P. A. Redhead, J. P. Hobson and E. V. Kornelsen, "The Physical Basis of Ultrahigh Vacuum," Chapman and Hall, 1968, p. 268.

[19] K. Jousten, in *Handbook of Vacuum Technology*, Wiley-VCH Verlag, 2008, p. 573.

[20] P. A. Redhead, J. P. Hobson and E. V. Kornelsen, in *The Physical Basis of Ultrahigh Vacuum*, Chapman and Hall, 1969, p. 267.

[21] V. Nemanic and J. Setina, "Outgassing in thin wall stainless steel cells," *J. Vac. Sci. Technol. A,* vol. 17, no. 3, pp. 1040-1046, 1999.

[22] K. M. Welch, Capture pumping technology, 2 ed., North-Holland, 2001.

[23] E. Fischer and H. Mommsen, "Monte Carlo computations on molecular flow in pumping speed test domes," *Vacuum,* vol. 17, no. 6, p. 309, 1967.

[24] P. A. Redhead, J. P. Hobson and E. V. Kornelsen, The Physical Basis of Ultreahigh Vacuum, Chapman and Hall, 1968, p. 275.

[25] K. Teraka, T. Okano and Y. Tuzi, "Conductance modulation method for the measurement of the pumping speed and outgassing rate of pumps in ultrahigh vacuum," *J. Vac. Sci. Technol. A,* vol. 7, no. 3, p. 2397, 1989.

[26] K. Saito, Y. Sato, S. Inayoshi and S. Tsukahara, "Measurement system for low outgassing materials by switching between two pumping paths," *Vacuum,* vol. 47, no. 6-8, p. 749, 1996.

[27] Y. Yang, K. Saito and S. Tsukahara, "An improved throughput method for the measurement of outgassing rates of materials," *Vacuum,* vol. 46, no. 12, p. 1371, 1995.

[28] H. F. Dylla, D. M. Manos and P. H. LaMarche, "Correlation of outgassing of stainless steel and aluminium with various surface treatments," *J. Vac. Sci. Technol.,* vol. 11, no. 5, pp. 2623-2636, 1993.

[29] D. Edwards Jr, "An upper bound to the outgasing rate of metal surfaces," *J. Vac. Sci. Technol.,* vol. 14, no. 4, pp. 1030-1032, 1977.

[30] G. F. Cerofolini, "Colloidal Science Volume 4," Royal Society of Chemistry, 1983, p. 79.

[31] G. F. Cerofolini, "Heterogeneity Effects in Desorptin Kinetics," *Z. phys. Chemie, Leipzig,* vol. 259, no. 6, pp. 1020-1024, 1978.

[32] G. F. Cerofolini, "Kinetics of Desorption from Heterogeneous Surfaces," *Langmuir,* vol. 13, pp. 990-994, 1997.

[33] K. Saito, Y. Sato, S. Inayoshi and S. Tsukahara, "Measurement system for low outgassing materials by switching between two pumping paths," *Vacuum,* vol. 47, no. 6-8, pp. 749-752, 1996.

[34] R. Renzi, Water Vapour in Metallic Vacuum Systems: Modelling and Experimental Studies for the LHC Injector Chain, Master Thesis, University of Padova (Italy), 2014.

[35] R. Weiss, H. F. Dylla and M. Zucker, "NSF Workshop on Large Ultrahigh-Vacuum Systems for Frontier Scientific Research," in *LIGO Document P1900072-v1*, https://dcc.ligo.org/public/0158/P1900072/001/P1900072-v1.pdf, 2019.

[36] R. E. Honig and H. O. Hook, *RCA Review,* vol. 21, pp. 360-368, 1960.

[37] K. Tatenuma, T. Momose and H. Ishimaru, "Quick acquisition of clean ultrahigh vacuum by chemical process technology," *J. Vac. Sci. Technol. A,* vol. 11, no. 4, pp. 1719-1724, 1993.

[38] S. R. Koebley and R. A. Outlaw, "Degassing a vacuum system with in-situ UV radiation," *J. Vac. Sci. Technol. A,* vol. 30, no. 6, pp. 060601-1, 2012.

[39] J. R. Vig, "UV/ozone cleaning of surfaces," *J. Vac. Sci. Technol. A,* vol. 3, p. 1027, 1985.

[40] P. W. Atkins and J. De Paula, Atkins' Physical Chemistry, 9 ed., p. 801.

[41] D. G. Bills, "Ultimate Pressure Limitations," *J. Vac. Sci. Technol.,* vol. 6, pp. 166-173, 1969.

[42] K. Kanazawa, "Analysis of pumping down process," *J. Vac. Sci. Technol. A,* vol. 7, no. 6, pp. 3361-3370, 1989.

[43] C. H. P. Lupis, Chemical Thermodynamics of Materials, North-Holland, 1983.

[44] S. Stolen and T. Grande, Chemical Thermodynamics of Materials: Macroscopic and Microscopic Aspects, John Wiley & Sons, 2004, p. 68.

[45] P. W. Atkins and J. De Paula, Atkins' Physical Chemistry, 9 ed., 2010, p. 890.

[46] T. Panczyc and W. Rudzinski, "Phenomenological Kinetics of Real Fas-Adsorption-Systems: Isothemal Kinetics and Kinetics of Thermodesorption," *J. Non-Equilib. Thermodyn.,* vol. 28, pp. 341-397, 2003.

[47] P. A. Redhead, "Modeling the pumpdown of a reversibly adsorbed phase: monolayer and submonolayer initial coverage," *J. Vac. Sci. Technol. A,* vol. 13, no. 2, pp. 467-475, 1995.

[48] B. B. Dayton, "Relations between size of vacuum chamber, outgassing rate, and required pumping speed," in *6th National Symposium on Vacuum Technology Transaction*, 1959.

[49] T. Krauss, "Uber die Evakuierungsgeschwindigkeit von Hochvakuumanlagen," *Vakuum Technik,* vol. 8, no. 2, pp. 39-43, 1959.

[50] A. Schram, "La désorption sous vide," *Le Vide,* vol. 103, pp. 55-68.

[51] P. W. Atkins and J. De Paula, Atkins' Physical Chemistry, Oxford University Press, 2010, p 894.

[52] W. Rudzinski and D. H. Everett, Adsorption of gases on heterogeneous surfaces, Elesevier Science and Technology, 1991.

[53] J. P. Hobson and R. A. Armstrong, *J. Phys. Chem.,* vol. 67, p. 2000, 1963.

[54] J. P. Hobson, "Theoretical submonolayer adsorption isotherms for hydrogen on a heterogeneous surface," *J. Vac. Sci. Technol. A,* vol. 13, no. 3, p. 549, 1995.

[55] J. P. Hobson, "Calculated physical adsorption isotherms of neon and radon on a heterogeneous surface," *J. Vac. Sci. Technol. A,* vol. 16, no. 3, p. 728+730, 1997.

[56] C. Benvenuti, J. P. Bojon, P. Chiggiato and G. Losch, "Ultimate pressures of the large electron positron collider (LEP) vacuum system," *Vacuum,* vol. 44, no. 5-7, pp. 507-509, 1993.

[57] H. Wipf , "Solubility and diffusion of hydrogen in pure metals and alloys," *Physica Scripta,* vol. T94, pp. 43-51, 2001.

[58] P. Tison, "Influence de l'hydrogène sur le comportement des metaux," Commissariat à l'Enérgie Atomique (CEA), Rapport CEA-R-5240(1), 1984.

[59] P. C. Novelli, P. M. Lang, K. A. Masarie, D. F. Hurst, R. Myers and W. Elkins, "Molecular hydrogen in the tropophere: Global distribution and budget," *J. Geophys. Res.,* vol. 104, no. D23, pp. 30427-30444, 1999.

[60] M. I. Baskes, "A calculation of the surface recombination rate constant for hydrogen isotopes in metals," *J. Nucl. Mat.,* vol. 92, pp. 318-324, 1980.

[61] J. Crank, The mathematics of diffusion, Second ed., Oxford Science Publications.

[62] G. Lewin, "Reduction of the Outgassing of Thick Stainless-Steel Sheets in Ultrahigh Vacuum by High-Temperature Vacuum Treatment," *J. Vac. Sci. Technol.,* vol. 6, pp. 420-423, 1969.

[63] M. R. Louthan and R. G. Derrick, "Hydrogen transport in austenitic stainless steel," *Corrosion Science,* vol. 15, no. 6-12, pp. 565-577, 1975.

[64] D. W. Hahn and M. Necati Özisik, Heat Conduction, Third ed., John Wiley & Sons, p. 120.

[65] "https://en.wikipedia.org/wiki/Fourier_number," [Online].

[66] P. Chiggiato, A. Harrison, M. Morrone, C. Pasquino and I. Wevers, "Reduction of hydrogen outgassing of austenitic stainless steel by thermal treatments: a comparative study between vacuum firing and air bakeout," *to be submitted to J. Vac. Sci. Technol.*.

[67] B. M. Shipilevsky and V. G. Glebovsky, "Competition of bulk and surface processes in the kinetiks of hydrogen and nitrogen evolution from metals into vacuum," *Surface Science,* vol. 216, pp. 509-527, 1989.

[68] R. L. Samuel, "Bake-out for UHV chambers in air at atmospheric pressure," *Vacuum,* vol. 20, no. 7, pp. 295-296, 1970.

[69] J. R. Young, "Outgassing characteristics of stainless steel and aluminium with different surface treatments," *J. Vac. Sci. Technol.,* vol. 6, pp. 398-400, 1969.

[70] Y. Ishikawa and T. Yoshimura, "Importance of the surface oxide layer in the reduction of outgassing from stainless steels," *J. Vac. Sci. Technol. A,* vol. 13, no. 4, pp. 1847-1852, 1995.

[71] M. e. a. Bernardini, "Air bake-out to reduce hydrogen outgassing from stainless steel," *J. Vac. Sci. Technol. A,* vol. 16, no. 1, pp. 188-193, 1998.

[72] V. Brisson, J. Hoang, G. Lissilour, P. Marin, A. Reboux, M. Bernardini and A. Pasqualetti, "Ultra-high vacuum qualification of the prototype module for the 2x3 km arms of the VIRGO interferometer," *Vacuum,* vol. 60, no. 1-2, pp. 9-14, 2001.

[73] A. Charlesby, Atomic radiation and polymers, Pergamon Press, 1960, p. 201.

[74] P. Beynel, P. Maier and H. Schönbacher, "Compilation of radiation damage test data. Part III: materials used around high-energy accelerators," CERN 82-10, 1982.

[75] J. C. Anderson, K. D. Leaver, P. Leevers and R. D. Rawlings, Materials Science for Engineers, 5 ed., CRC Press - Taylor & Francis Group, 2003, p. 380.

[76] S. Kavesh and J. M. Schultz, "Meaning and Measurement of Crystallinity in Polymers: A Review," *Polymer Engieneering and Science,* vol. 9, no. 5, pp. 331-338, 1969.

[77] A. S. Michaels, W. R. Vieth and J. A. Barrie, "Solution of Gases in Polyethylene Terephthalate," *Journal of Applied Physics,* vol. 34, no. 1, pp. 1-12, 1963.

[78] A. S. Michaels, W. R. Vieth and J. A. Barrie, "Diffusion of Gases in Polyethylene Terephthalate," *Journal of Applied Physics,* vol. 34, no. 1, pp. 13-20, 1963.

[79] J. Guo and T. A. Barbari, "Unified dual mode description of small molecules sorption and desorption kinetics in a glassy polymer," *Macromolecules,* vol. 42, pp. 5700-5708, 2009.

[80] S. G. Sammartano, "Outgassing rates of PEEK, Kapton and Vespel polymers," Master Thesis, Arcada (Helsinki, Finland), 2020.

[81] E. H. H. Jamieson and A. H. Windle, "Structure and oxygen-barrier properties of metallized polymer film," *Journal of Materials Science,* vol. 18, pp. 64-80, 1983.

[82] H. Chatham, "Oxygen diffusion barrier properties of transparent oxide coatings on polymeric substrates," *Surface and Coatings Technology,* vol. 78, pp. 1-9, 1996.

[83] A. S. da Silva Sobrinho, M. Latrèche, G. Czeremuszkin, J. E. Klember-Sapieha and M. R. Wertheimer, "Transparent barrier coatings on polyethylene terephthalate by single- and dual-frequency plasma-enhanced chemical vapor deposition," *J. of Vac. Sci. & Technol. A,* vol. 16, no. 6, pp. 3190-3198, 1998.

[84] C. Bellachioma, "Applicazione di film sottili per il controllo della diffusione e della permeabilità di materiali polimerici per UHV," Università degli Studi di Perugia, PhD Thesis, CERN-THESIS-2004-060, http://cds.cern.ch/record/1379846, 2004.

[85] H. Prins and J. J. Hermans, "Theory of Permeation through Metal Coated Polymer Films," *J.Phys.Chem.,* vol. 63, no. 5, pp. 716-720, 1959.